\documentclass{article}
\usepackage{authblk}
\usepackage{lineno,hyperref}

\usepackage{amssymb}
\usepackage{latexsym}

\usepackage{multirow}
\usepackage{epsfig}
\usepackage{graphicx}
\usepackage{amsmath}
\usepackage{rotating}
\usepackage[version=4]{mhchem}
\usepackage{listings} 
\usepackage[utf8]{inputenc}
\usepackage[english]{babel}
\usepackage{verbatim}
\usepackage{color}
\usepackage{dsfont}
\usepackage{subfig}
\usepackage{lastpage}			
\usepackage{indentfirst}		
\usepackage{color}				
\usepackage{graphicx}			
\usepackage{microtype} 			
\usepackage{amsmath}
\usepackage{amssymb}
\usepackage{latexsym}
\usepackage{epsfig}
\usepackage{rotating}
\usepackage{tabularx,ragged2e}
\usepackage{tabu}
\usepackage{appendix}
\usepackage{subfig}
\usepackage{rotating}
\usepackage{tabto}
\usepackage[table]{xcolor}
\usepackage{bm}
\usepackage{epigraph}
\usepackage{multirow}
\usepackage[version=4]{mhchem}
\usepackage{float}
\usepackage{makecell}
\usepackage{adjustbox}
\usepackage{newfloat}
\usepackage{verbatim}
\usepackage{ulem}
\usepackage{tikz}
\usetikzlibrary{arrows,positioning}
\usetikzlibrary{calc}
\usetikzlibrary{decorations.pathreplacing}
\usepackage{ctable}
\usepackage{framed}

\usepackage{newfloat}
\DeclareFloatingEnvironment[
  fileext = los ,
  listname = {List of Schemes} ,
  name = Scheme
]{schemea}

\newcommand{\be}{\begin{eqnarray}}
\newcommand{\ee}{\end{eqnarray}}
\newcommand{\ba}{\begin{array}}
\newcommand{\ea}{\end{array}}
\newcommand{\bmn}{\begin{minipage}}
\newcommand{\emn}{\end{minipage}}
\newcommand{\bv}{\begin{verbatim}}
\newcommand{\ev}{\end{verbatim}}
\newcommand{\bt}{\begin{tabular}}
\newcommand{\et}{\end{tabular}}
\newcommand{\bi}{\begin{itemize}}
\newcommand{\ei}{\end{itemize}}
\newcommand{\btab}{\begin{table}}
\newcommand{\etab}{\end{table}}
\newcommand{\btabs}{\begin{table*}}
\newcommand{\etabs}{\end{table*}}
\newcommand{\btabsw}{\begin{sidewaystable}}
\newcommand{\etabsw}{\end{sidewaystable}}
\newcommand{\bfig}{\begin{figure}}
\newcommand{\efig}{\end{figure}}
\newcommand{\bfigs}{\begin{figure*}}
\newcommand{\efigs}{\end{figure*}}
\newcommand{\bc}{\begin{center}}
\newcommand{\ec}{\end{center}}
\newcommand{\bit}{\begin{itemize}}
\newcommand{\eit}{\end{itemize}}

\definecolor{darkgreen}{rgb}{0.0, 0.3, 0.0}

\title{Metabolic Network Properties: Comprehensive Analysis Across Domains}

\author[1]{José Antônio Pellizzaro}
\author[1]{Daniel Gamermann\thanks{danielg@if.ufrgs.br}}
\author[2]{Julian Triana Dopico}
\affil[1]{Department of Physics, Universidade Federal do Rio Grande do Sul (UFRGS) - Instituto de Física \\ Av. Bento Gonçalves 9500 - Caixa Postal 15051 - CEP 91501-970 - Porto Alegre, RS, Brasil.}
\affil[2]{Center for Regenerative Biotherapeutics, Mayo Clinic Hospital, Phoenix, AZ, United States of America}

\date{\today}

\begin{document}

\maketitle 
\begin{abstract}
Metabolic networks play pivotal roles in understanding the evolution of organisms, microbiome dynamics and disease prevention and treatment. This study
presents a comprehensive analysis of metabolic network properties across 10912 organisms spanning Bacteria, Archaea, and Eukarya domains. A novel method
for {the} network construction is introduced, emphasizing the chemical transformations of metabolites. Unlike conventional approaches that link {every} substrate to every product {in all chemical reactions}, this method aligns more closely with metabolic pathways, linking products only to their generating substrates. Emphasis is placed on investigating the community structure within metabolic networks, employing the Surprise quality function {that better} addresses {some} limitations of previous approaches. Through comparisons between real and randomized {versions of the} networks, we identify characteristics that cannot be explained solely by the network degree distribution{, thus identifying} topological properties and community structures {that} arise from additional evolutionary pressures. This highlights the need for evolutionary models to incorporate additional mechanisms beyond the replication of the degree distribution.
\end{abstract}




\section{Introduction}

The study of metabolic networks is relevant in various fields. These graphs may be used to illuminate the evolution of organisms \cite{kreimer2008evolution, zhou2012convergent, basler2012evolutionary, ramon2023functional}, understand the structure and functioning of microbiomes \cite{muller2018using, cardona2016network}, or explore the treatment and prevention of diseases by identifying potential drug targets \cite{rahman2006observing, yeh2004computational}. 
In this context, our work constitutes a comprehensive study of the properties of metabolic networks. We conducted an analysis of the metabolic networks of 10912 different organisms, classified into three taxonomic domains: Bacteria, Archaea, and Eukarya.

A key aspect of our work is the introduction of a novel method for {the} network creation that better encapsulates the transformation of each {metabolite (compound, molecule, element, etc.)} in the network. Unlike the standard practice in the literature, where each substrate is linked to each product {in every chemical reaction} \cite{zhou2012convergent, takemoto2014metabolic,ravasz2002hierarchical, ma2003reconstruction, takemoto2007correlation, gamermann2}, our method {first} considers the chemical composition of each metabolite. Specifically, metabolites are linked only to compounds that directly contribute to their formation. This representation is more closely aligned with the concept of a metabolic pathway, allowing a clearer reconstruction of how metabolites are synthesized (anabolism) or broken down into simpler products (catabolism).

In addition, special emphasis will be given to the study of the community structure of metabolic networks. Although there is no consensus on a formal definition, communities are derived from the intuitive idea that nodes form subgroups within the larger network. These groups have particular interest in the field of molecular biology as{, for example,} they are associated with specific biological functions \cite{hartwell1999molecular}. Previous studies in this area focused on detecting communities by maximizing the Modularity function \cite{kreimer2008evolution, zhou2012convergent,  ravasz2002hierarchical, good2010performance, koch2013functional}. However, it has been shown that this function has significant drawbacks, including a resolution limit and a degeneracy problem \cite{good2010performance, Fortunato36, nicolini2017community,nicolini2016modular, gamermann2022algorithm}. Due to these limitations, we have chosen to use a different quality function, the Surprise \cite{arnau2004iterative}, which does not suffer from the same drawbacks as Modularity {does} \cite{gamermann2022algorithm, aldecoa2011deciphering, traag2015detecting}.

We also investigate whether the topological properties of the networks and the characteristics of their community structure are {solely} a direct consequence of the degree distribution of the vertices in the graph or whether they result from different evolutionary pressures. To assess this, we created randomized versions of each metabolic network, ensuring that they share the same degree distribution as their real counterparts. By comparing the properties of the randomized networks with those of the real ones, we can evaluate the statistical significance of any observed differences. Consequently, this approach allows us to identify attributes that are not mere consequences of the degree distribution of the network. This is relevant because many evolutionary models of biological networks \cite{vazquez2003growing,krapivsky2001organization,albert2002statistical, barabasi1999emergence, knight} have been designed primarily to reproduce the power-law degree distributions, while paying less attention to other topological properties. If {one} can attest that certain properties are not a direct result of the degree distribution {alone}, then these evolutionary models should include other mechanisms to account for the differences. 

This work is organized as follows: in Section \ref{sec:methods}, we introduce our novel network construction method and present the sources of metabolic network and taxonomic data. Next, in section \ref{sec:ana} we describe the analysis done and the randomization process we used. In section \ref{sec:results} we present our results and finally, in section \ref{sec:conc} we do a brief overview and draw our conclusions. An appendix is also included, where one finds details and examples on how the metabolic networks are constructed by linking substrates and products in the chemical reactions, based on their composition similarity.

\section{Materials and Methods}\label{sec:methods}

\subsection{Network Construction}

The metabolism of an organism is commonly depicted as a substrate network \cite{takemoto2014metabolic,ravasz2002hierarchical, ma2003reconstruction, takemoto2007correlation, gamermann2}. In these networks, each node represents a metabolite, and nodes are connected if they constitute a substrate-product pair in a biochemical reaction within the metabolism. It is important to note that some works \cite{wang, schnel, takemoto2012current} have already pointed out drawbacks with this simple representation.

In the present work, we improve upon the substrate network representation. Instead of linking each substrate to each product {(molecules on the left-hand side of a reaction to the molecules on the right-hand side)}, we want to determine how each individual molecule changes through {a sequence of} chemical reactions. {In many reactions, some of the metabolites participate primarily as donors or acceptors of energy, electrons, or functional groups. Examples include ATP/ADP and cythocromes. Although these compounds participate in the reaction, they are not directly involved in the chemical transformation of the main metabolites. Consequently, a pathway-based interpretation does not always justify connecting them to every product and substrate in a reaction. In other words, {for the construction of our networks,} for each biochemical reaction, we want to answer the question: what gets transformed into what? 

In order to create the metabolic network of an organism, we need the list of every biochemical reaction present in its metabolism. This information was acquired from the Kyoto Encyclopedia of Genes and Genomes (KEGG) database \cite{kegg}. {For each organism, a list of enzymes identified in its genome is obtained.} Using this list of enzymes, one extracts the catalog of the biochemical reactions catalyzed by each enzyme. With the list of reactions, one can finally build the networks.

Here, we are not interested in the shape of molecules {(stereochemistry), but} only in their chemical compositions as we seek to reconstruct each {metabolite by grouping the different molecules that combined to form it or into which it broke down.} {Please, refer to the Appendix \ref{appendixA} for a detailed explanation on how the algorithm runs.}

In KEGG's database, each reaction, compound, and glycan is represented by a code: RXXXXX, CXXXXX, and GXXXXX, respectively. In each case, the X's symbolize a set of integers that uniquely define each reaction or molecule. For instance, KEGG’s reaction R05740 is stored as
\begin{center}
    
    \ce{\text{C01595} + \text{2 C00999} + \text{C00007} + \text{2 C00080} <=> \text{C07289} + \text{2 C00996} + \text{2 C00001.}}

\end{center}
Substituting the codes for the compound names and chemical formulas gives {the following}:
\begin{align*}
    \ce{$\underbrace{\textrm{Linoleate}}_{C_{18}H_{32}O_{2}} + 2 \underbrace{\textrm{Ferrocytochrome b5}}_{Ferrocytochrome b5_{}} + \underbrace{\textrm{Oxygen}}_{O_{2}} + 2 \underbrace{\textrm{H+}}_{H_{}} \leftrightarrow \underbrace{\textrm{Crepenynate}}_{C_{18}H_{30}O_{2}} + 2 \underbrace{\textrm{Ferricytochrome b5}}_{Ferricytochrome b5_{}} + 2 \underbrace{\textrm{H2O}}_{H_{2}O_{}}$}.
\label{react:1}
\end{align*}

Our algorithm will {first} connect Linoleate with Crepenyate {then} hydron (\ce{H+}) and oxygen with water and finally, since it was unable to produce a chemical formula for Ferricytochrome b5 and Ferrocytochrome b5, will connect one to the other. Figures \ref{fig:net_react_1}-a) and b) show the comparison between the substrate networks found in the literature, where {one} simply connects each {compound in the left-hand side with each one in the right-hand side}, and our new method, where we link the molecules according to their chemical compositions. {Note that in this reaction the cytochromes are only fulfilling the role of electrons donor or acceptor and do not contribute to the the chemical (structure) transformation of the other molecules.}

\begin{figure}[htbp]
\makebox[\linewidth][c]{
    \centering

    \subfloat[Traditional Substrate Network]{\includegraphics[width=0.5\textwidth]{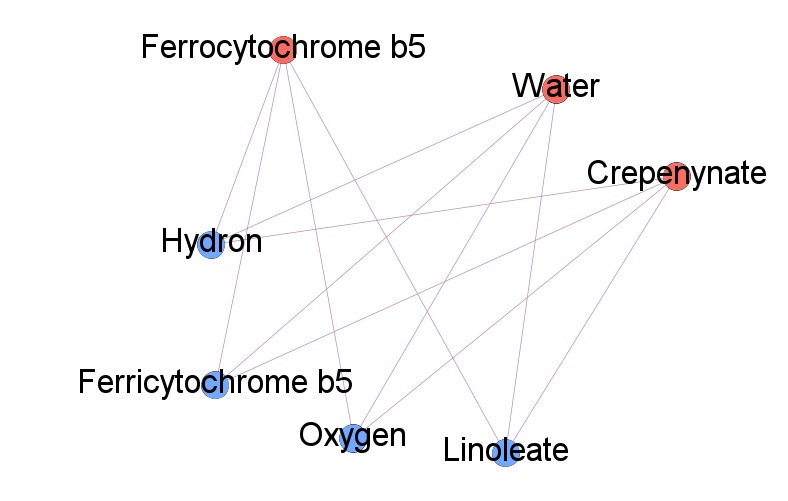}}    
   \subfloat[New Network Representation]{\includegraphics[width=0.5\textwidth]{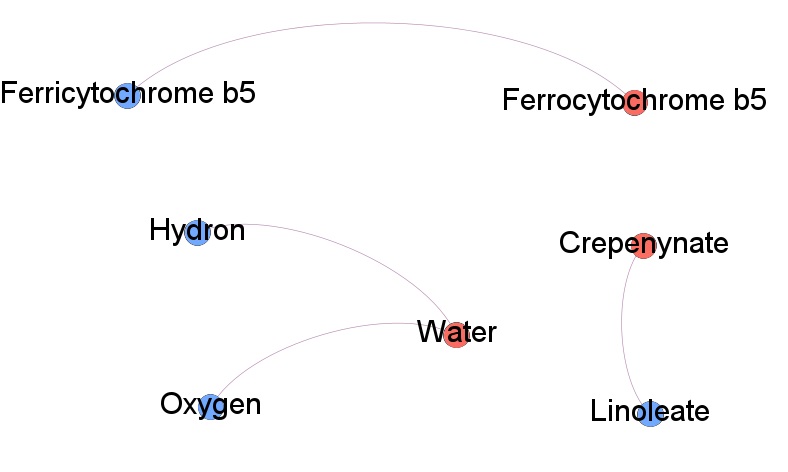}}
}
    \caption{In a) and b) we have the visual representation of the links from reaction R05740 using our old method (a) and our new method (b). In a) we simply link each product in red to each substrate in blue, while in b) we link the molecules based on their chemical composition.}
    \label{fig:net_react_1}
\end{figure}

Although a comprehensive biological validation of every reaction assignment is beyond the scope of this work, the proposed method is grounded on elemental balance and molecular composition, ensuring that edges represent chemically meaningful transformations rather than generic co-participation in reactions. Our representation has two main advantages. First, it better captures the definition of a metabolic pathway, where a metabolite is transformed by a chain of reactions (in a pathway, the product of one reaction is the substrate for subsequent reactions). 

Second, it indirectly solves the issue of currency metabolites \cite{takemoto2014metabolic,ravasz2002hierarchical, ma2003reconstruction}. Currency metabolites, such as water, ATP, NADH, oxygen, cytochromes, or other specific peptides, and some RNAs, participate in the exchange of energy, electrons, protons, phosphate moieties, or perform other secondary functions and are ubiquitous in metabolisms \cite{takemoto2014metabolic}. Their indiscriminate inclusion can substantially alter the network topology. In particular, currency metabolites create highly connected hubs that artificially shorten the distance between otherwise unrelated metabolites \cite{ma2003reconstruction}. Therefore, a common practice in the literature is to simply exclude these currency metabolites from the network \cite{takemoto2014metabolic, ravasz2002hierarchical, takemoto2007correlation}. However, this approach has limitations because the classification of a metabolite as currency is context-dependent and in some reactions, molecules such as water, oxygen {and even ATP or ADP} can be primary metabolites.
 
Our method bypasses this problem. By constructing the products of biochemical reactions as combinations of their substrates, we identify the currency metabolites automatically. Returning to the example in Fig. \ref{fig:net_react_1}, if each substrate were simply connected to each product, we would erroneously link water to Linoleate and Ferricytochrome b5, even though this connection lacks a biochemical sense. Our algorithm accurately identifies water (and, by extension, \ce{H^+} and \ce{O_2}) as currency metabolites and appropriately links them to each other. The primary relevant chemical transformation in this reaction, for the organism, is the transformation of Linoleate to Crepenyate (or vice-versa).

The primary challenge in our approach was to determine the chemical formula of glycans. Glycans have compositions where each element can be a huge molelule. We followed the Nomenclature for Glycans (SNFG)\cite{varki2015symbol} and tried to resolve the chemical formula of a glycan from its components by sistematically searching and identifying them in the database.


\section{{Network Properties}}\label{sec:ana}

In this section, we define the attributes evaluated for each {one} of the 10912 networks created\footnote{{Initially we created the networks for 10950 organisms, but then realized some were referring to different strains or subspecies of the same organism.} We {then} employed the following criteria for selecting only one: first, we utilized the reference genome when available; second, we prioritized genomes with references from the literature. Whenever this information was unavailable, we selected the most recent genome for that species. {Ultimately, our analysis involved 10,912 organisms, including 9273
bacteria, 448 archea and 1191 eukaryotes.}} and their randomized versions. Given our interest in comparing metabolic networks across the three domains of life, we focus on global and community-level properties rather than local (node-level) ones.

Since we are studying undirected and unweighted graphs, each network is represented by a square \( N \times N \) adjacency matrix \( A \), where \( N \) is the number of nodes in the network. In this matrix, \( A_{ij} = 1 \) if there is a link between nodes \( i \) and \( j \), and \( A_{ij} = 0 \) otherwise. 

In addition to basic properties such as the number of nodes $N$ and the number of links $L$, we also evaluate the network degree distribution, its clustering coefficients and different node correlation metrics: their assortativity and the {average} shortest path distance between nodes. {We present in Table \ref{tab:eqs} the main attributes that will be analyzed for each of the graphs studied in this paper: the parameter name, mathematical definition and a brief description.}

The degree of a node $k_i$ is equal to the number of links it has in the network. The clustering coefficient $C_i$ {of node $i$ is a local parameter that} measures the extent to which its neighbors are interconnected. Averaging these values allows us to obtain network-level properties: the average degree of the network $\bar{k}$ and its average local clustering coefficient $\bar{C}$. However, it is also possible to calculate the network's global clustering coefficient $\mathcal{C}$, by comparing the ratio of 3-cliques (the number of triangles $C_3$) with the number of connected triples (2-paths, $P_2$), as described in Table \ref{tab:eqs}. Later, we will demonstrate that the values of $\mathcal{C}$ and $\bar{C}$ are significantly different in metabolic networks.

At the network level, we explored whether its degree distribution followed a power law ($p(k) \propto k^{-\gamma}$), a property often called scale-freeness. Although traditionally considered prevalent in real world networks, recent studies have raised doubts on this notion \cite{gamermann2, broido2019scale, khanin2006scale}.
We investigated this by fitting a discrete power-law model to the degree distribution of each network in the dataset. A detailed explanation of the fitting process is given in a previous study \cite{gamermann2}. The goodness of fit is evaluated using a $\chi^2$ test where a high p-value indicates a good fit (high probability of making a type-I error if the power-law hypothesis is discarded).

We also examine the assortativity ($-1\le A\le 1$), which measures the correlation between the degrees of connected nodes. In an assortative network ($A>0$), high-degree nodes are connected to other high-degree nodes, whereas in disassortative networks ($A<0$), high-degree nodes {tend to be} connected to low-degree ones. {An $A$ close to zero means no tendency on the matter.}

Dijkstra's algorithm \cite{dijkstra2022note} was employed to calculate the shortest path distance between each pair of nodes in {a} network. These distances populate the symmetric distance matrix, where each element \( d_{ij} \) represents the shortest path length between nodes \( i \) and \( j \). The network's average distance \( \bar{d} \) is then determined by averaging all elements in the {upper triangle of the} distance matrix.

Finally, since most of the constructed networks had small disconnected components, we also count, for each of them, the number of disconnected components, the size of the largest component, and the average size of the smaller components\footnote{These smaller components, as we will discuss later, are usually pairs like the Ferro- and Ferri- cytochromes, shown in the previous reaction example that, though complex structures, they do not breakdown or come to be as new molecules merged in the reactions pertaining to the metabolism {\it perse}. They are, instead, some sort molecular ``gear'' the reaction needs for the transformation to take place.}.

\begin{table}[h]
\centering
\begin{adjustbox}{max width =\textwidth}
\begin{tabular}{p{4cm}c|c|p{6cm}l}
\hline
Parameter & Math Symbol & Equation & Description \\ \hline
& & &  \\
Degree & \( k_i \) & \( k_i = \sum\limits_{j=1}\limits^N A_{ij} \) & Number of connections a node has. \\
& & &  \\
Average Degree & \( \bar{k} \) & \( \bar{k} = \frac{1}{N} \sum\limits_{i=1}\limits^N k_i \) & Average number of connections per node. \\ 
& & &  \\
Local Clustering Coefficient & \( C_i \) & \( C_i = \frac{2E_i}{k_i(k_i-1)} \) & Ratio of existing triangles to possible triangles among neighbors. Here $E_i$ represents the number of links between the neighbors of node $i$. \\ 
& & &  \\
Average Local Clustering Coefficient & \( \bar{C} \) & \( \bar{C} = \frac{1}{N} \sum\limits_{i=1}\limits^N C_i \) & Average clustering coefficient of all nodes. \\ 
& & &  \\
Global Clustering Coefficient & \( \mathcal{C} \) & \( \mathcal{C} = 3\frac{C_3}{P_2} \) & Overall measure of clustering in the network. \\ 
& & &  \\
Average Distance & \( \bar{d} \) & \( \bar{d} = \frac{2}{N(N-1)} \sum\limits_{i=1}\limits^N \sum\limits_{j>i} d_{ij} \) & Mean shortest path distance between all pairs of nodes. \\ 
& & &  \\
Assortativity & \( A \) & $ A = \frac{\frac{1}{L}\sum_l k_i(l)k_j(l) - \left[\frac{1}{L}\sum_l\left(k_i(l) + k_j(l)\right)\right]^2}{\frac{1}{L}\sum_l\left(k_i^2(l) + k_j^2(l)\right) - \left[\frac{1}{L}\sum_l\left(k_i(l) + k_j(l)\right)\right]^2} $
&  Pearson correlation coefficient of the degrees of nodes $k_i(l)$ and  $k_j(l)$ at the ends of link $l$. \\ 
& & &  \\
{Maximum possible number of links in the network} & $F$ & $F = \frac{N(N-1)}{2}$ & {Parameter of the Surprise function. Equivalent to the number of links in a clique of size $N$.} \\
& & & \\
{Maximum number of intracommunity links in the partition} & $M$ & $\displaystyle M = \sum_{i=1}^{N_c} \frac{c_i(c_i -1)}{2}$ &  {Parameter of the Surprise function. We consider that the network is partitioned into $N_c$ communities where the size of community $i$ is $c_i$.} \\
& & & \\
{Surprise} & \( S \) & $\displaystyle S = - \log \sum_{j=\zeta}^{\textrm{min}(M,L)} \frac{\binom{M}{j} \binom{F-M}{L-j}}{\binom{F}{L}}$. & {An accumulated hypergeometric distribution that measures how surprising (unlikely) it is to find
a partition with as many intracommunity links as in the given graph. The parameters $F$ and $M$ are described above, while $L$ and $\zeta$ represent the total number of links in the network and the number of intracommunity links in the partition, respectively.}
 \\
& & & \\
\hline
\end{tabular}
\end{adjustbox}
\caption{Summary of network properties and their mathematical definitions.}
\label{tab:eqs}
\end{table}

\subsection{Community Structure}

We obtained the community structure of the networks using the Surpriser algorithm \cite{gamermann2022algorithm}. This algorithm aims to {find the partition that} maximizes the value of the Surprise function \cite{gamermann2022algorithm,arnau2004iterative, aldecoa2011deciphering, traag2015detecting}. The Surprise is an accumulated hypergeometric distribution that measures how surprising (unlikely) it is to find a partition with as many intracommunity links as in the given graph. 

{The formulation of the Surprise (see Table \ref{tab:eqs}) is equivalent to an urn problem. In this analogy, every \textit{possible link} $F$ corresponds to a ball in the urn: $(F - M)$ of them are red and represent failures (extra-community links), while $M$ are green and represent successes (intra-community links). For each partition, the proportions of red and green balls vary because $M$ depends on the community structure. Surprise measures the probability of obtaining at least $\zeta$ successes (the number of intra-community links in the partition) in $L$ draws (as there are $L$ links in the actual network) without replacement.}

{Alongside the parameters of the Surprise function, $M$, $\zeta$, and the Surprise value $S$ itself, we also analyzed the distribution of community sizes. In particular, we examined the number of communities $N_c$, the average community size $\bar{n}_c$, the size of the largest community $\max(n_c)$, and the evenness of {the} community size distribution through its Pielou’s evenness index, $PI$.}

\subsection{Randomization Process}

In this work, we follow the randomization process described in a previous study \cite{gamermann2}. Our goal was to create randomized versions of {the} metabolic networks that share the same degree distribution as their real counterparts. 

{The method to construct such a network is quite straightforward: at each step, we randomly select two links from the network and break them; then, the nodes participating in the first link are connected to the nodes participating in the second link. We repeat this process enough times to ensure that each link has a 99.9\% chance of being altered at least once in the randomization process.} {Mathematically, given the space of all graphs with fixed number of nodes, $N$, fixed number of links, $L$, and a fixed distribution for its $N$ nodes degrees, $k_i$, this process would be the equivalent of randomly sampling the entire space of such graphs, thus} simulating the results of an evolutionary model constructed to generate networks that adjust the degree distribution observed in real-world graphs, without considering any other aspect of the resulting graphs. By comparing the characteristics of these randomized networks with those of the real graphs, one can discern the characteristics that arise {solely} from the network's degree distribution from those that result from different mechanisms. To access the statistical significance of these differences, we used a ten-sample student's t-test. The p-value for the null hypothesis that the observed network has the parameter equal to its expected value given the degree distribution (average over the property in the random sample) is equal to two times the value of the cumulative student’s t-distribution with 9 degrees of freedom\footnote{{We created 10 randomized copies for each metabolic network, hence the choice of 9 degrees of freedom. Since our dataset comprises 10912 different metabolic networks, choosing a number much higher than
10 would greatly increase the computational costs associated with this work. As it stands, we analyzed
a total of 120032 networks, which includes the original 10912 networks and 10 randomized copies of
each.}} at the point $|t_p|$. The value of $t_p$ associated with each parameter $P$ is defined as:
\begin{equation}
    t_p = \frac{\bar{P}_{\textrm{random}} - P_{\textrm{real}}}{\frac{S}{\sqrt{10}}},
\end{equation}
where $P_{\textrm{real}}$ represents the value of parameter $P$ in the real network, $\bar{P}_{\textrm{random}}$ symbolizes the average of parameter $P$ in the 10 randomized networks and $S$ represents the standard deviation in the random samples. A high p-value implies that the distinction between the real and random networks is not significant, indicating a high likelihood of obtaining a fluctuation equal to or larger than the observed one in the population. This also suggests that the parameter in question is related to the network's degree distribution, as they are the same between the real metabolic networks and their randomized versions.
However, a small p-value indicates a meaningful deviation between the real network parameter and the expected value in the random samples. 

\section{Results and Discussion}\label{sec:results}

\subsection{Real Metabolic Networks}
In this section, we will examine the {parameters} described in Section \ref{sec:ana} (which we will call topological properties from now on) and the community structure from the metabolic networks. Table \ref{tab:metab_top} shows the average value, the standard deviation, and the skewness for each topological property, while figures \ref{fig:metab_top}-a) to f) show the histograms of such parameters.

\btab[htbp]
    \centering
    \begin{adjustbox}{max width=0.8\textwidth}
    \bt{p{4cm}c|ccc}
    \hline
    Parameter & Math symbol & Average & Standard deviation & Skewness \\
    \hline
     & & & & \\
    Number of nodes & $N$ & $1399.851540_{-479.032810}^{+467.377529}$ & $472.988281$ & $0.096334$ \\ 
  & & & & \\ 
Number of links & $L$ & $2208.235704_{-818.400733}^{+803.240828}$ & $810.573468$ & $0.083944$ \\ 
  & & & & \\ 
Average degree & $\bar{k}$ & $3.107625_{-0.271490}^{+0.126372}$ & $0.193626$ & $-2.028022$ \\ 
  & & & & \\ 
Average local clustering & $\bar{C}$ & $0.073213_{-0.017772}^{+0.010880}$ & $0.014060$ & $-1.060120$ \\ 
  & & & & \\ 
Global Clustering & $\mathcal{C}$ & $0.025406_{-0.004974}^{+0.010410}$ & $0.007455$ & $2.262596$ \\ 
  & & & & \\ 
Assortativity & $A$ & $-0.107364_{-0.011171}^{+0.035156}$ & $0.022042$ & $4.618415$ \\ 
  & & & & \\ 
Average distance & $\bar{d}$ & $4.141408_{-0.109659}^{+0.195144}$ & $0.150074$ & $1.471783$ \\ 
  & & & & \\ 
Number of components & $N_{comps}$ & $75.348515_{-21.225522}^{+23.005418}$ & $22.114545$ & $0.212889$ \\ 
  & & & & \\ 
Size of main component & ${\mathcal{S}}_{main}$ & $1241.947122_{-438.506644}^{+422.710390}$ & $430.328701$ & $0.070655$ \\ 
  & & & & \\ 
Average size of smaller components & $\mathcal{S}_{small}$ & $2.130004_{-0.051491}^{+0.091708}$ & $0.070594$ & $1.781649$ \\ 
  & & & & \\ 
\hline
    \et
    \end{adjustbox} 
 \caption{Topological properties of metabolic networks. The plus minus super and subscripts next to the averages are the standard deviations when considering only values above and below the mean value separately, which makes an uncertainty bar asymmetric for parameters with an important skewness.}
\label{tab:metab_top} 
\etab

We observe that all metabolic networks are disassortative ($A < 0$), indicating a tendency for high degree nodes to be connected with low degree nodes, but this is expected given the degree distribution of the networks, where many more nodes have low degrees and, as a consequence, high degree nodes must connect with low degree ones. Additionally, these metabolic networks exhibit a higher average local clustering coefficient, $\bar{C}$ compared to their global clustering coefficient, $\mathcal{C}$. This lack of correlation between local and global clustering coefficients in real-world networks has been observed in various systems \cite{newman2003structure, estrada2012structure}, aligning with our previous findings \cite{gamermann2}. There are, however, significant differences between the networks analyzed in this study and those in earlier works. In the previous study \cite{gamermann2}, we created the aforementioned substrate networks, disregarding the transformation that each molecule undergoes in a biochemical reaction. The comparison between these works is striking. Although the number of nodes in our current study is approximately 55.3\% higher than in \cite{gamermann2} (due, possibly, to new information in KEGG's database), the number of links is 16\% smaller. This suggests that many connections present in traditional substrate networks may not correspond to direct metabolite transformations. To underscore this point, the average degree has decreased by 54\%, from 5.69 to 3.11, indicating that the networks are less interconnected than previously thought. Another key difference is that the present networks exhibit higher values of average distance and assortativity. These changes can be attributed to our treatment of currency metabolites. Currency metabolites typically form large hubs in substrate networks, which reduce the average distance between nodes and decrease network assortativity by connecting many low-degree nodes to high-degree ones (the hubs themselves). By restricting these connections to only those with biological significance, we reduce both effects, leading to an increase in average distance and assortativity.

We should note that in real metabolic networks, one should not expect disconnected components. Table \ref{tab:metab_top} shows the number of components ($N_{\text{comps}}$), the size of the largest (the main) component ($\mathcal{S}_{\text{main}}$), and the average size of the smaller components ($\mathcal{S}_{\text{small}}$) found in the networks we created. We can observe, however, that the average size of the disconnected components {in a given organism is around 2, very small} compared to the size of the main component. When comparing the average number of nodes with the average size of the main component, we find that almost 90\% of the nodes are part of the main component (88.71\%). Inspection of these two-node components reveals that they frequently involve large molecular structures such as peptides or RNA molecules. These compounds often do not participate directly in metabolic transformations. Instead, they perform auxiliary functions, such as electron transfer, co-factor exchange, or participation in transcriptional and translational processes. This observation is consistent with the examples discussed previously. Other observed cases are common enzymes (e.g. EC 5.3.1.6) that beyond the common reaction it catalyzes, are also able to perform a similar transformation using a different, but rare, substrate. When this is the case, the enzyme will be associated with more than one reaction. One of them relevant for an organism, plus another reaction that is not and might never be actually taking place. Our algorithm just collects this set of reactions, and, as already stated, it is beyond our scope to manually check the relevance of each reaction identified. Therefore, some reactions might create such small communities with metabolites that are irrelevant to the organism as a whole. In Table \ref{tab:smallcomps} we show, as examples, the five most prevalent of these small disconnected components that appear in our networks.

\btab[htbp]
    \centering
    \begin{adjustbox}{max width=0.9\textwidth}
    \bt{cc|cp{9cm}}
    \hline
    Compounds & Names & Frequency & Explanation \\
    \hline
C01642, C02412 & tRNA(Gly), Glycyl-tRNA(Gly) & 10846 (99.40\%) & tRNA coupling to amino acid.\\  
C01650, C02553 & tRNA(Ser), L-Seryl-tRNA(Ser) & 10820 (99.16\%) & tRNA coupling to amino acid.\\ 
C00886, C01635 & L-Alanyl-tRNA, tRNA(Ala) & 10782 (98.81\%) & tRNA coupling to amino acid.\\ 
C02962, C18096 & D-Allose 6-phosphate, D-Allulose 6-phosphate & 10467 (95.92\%) & Rare alternative substrate-product pair of a common enzyme.  \\
C03633, C03798 & Peptidylproline (omega=0), Peptidylproline (omega=180) & 10442 (95.69\%) & Change of polarization in a peptide.\\
\hline
    \et
    \end{adjustbox} 
 \caption{Top five most prevalent disconnected components. The first two columns show the members of the components with their KEGG codes and names, while the third shows in how many organisms this component was found (the percentage it represents) and the last column explains what this community means.}
\label{tab:smallcomps} 
\etab

{Table \ref{tab:gamma} presents the results for the fitting procedure of the power-law function to the degree distribution of the nodes in each network. The value of $x_0$ denotes the minimum degree considered in the fitting process; for example, an $x_0$ of 3 indicates that only nodes with $k\geq3$ were included in the fit (the tail of the distribution). Conversely, the corresponding fraction of discarded nodes is shown in the third column. Our findings indicate that the vast majority of metabolic networks generated using our method do not follow a power-law distribution. Of more than 160,000 fitting attempts, only 647 cases (approximately 0.40\%) yielded statistically significant results (\( p \geq 0.05 \)). A clear pattern emerges from these rare cases: networks that fit a power-law tend to be significantly smaller. On average, networks that exhibit a good fit contain only \( 344.69 \pm 144.20 \) nodes, compared to the average for the entire dataset of 1399 nodes. Moreover, in 75.12\% of these well-fitted cases, more than 90\% of the nodes were discarded during the fitting process. This suggests that the apparent power-law behavior arises primarily from the small network size or the limited number of nodes included in the analysis, rather than being an inherent structural property of the metabolic networks. To further investigate the degree distribution on a larger scale and assess whether any global trends emerge, we computed the Cumulative Distribution Function (CDF), \( P(x \geq x_0) \), for the combined data of all organisms. The plot of this distribution in logarithmic scale, along with the CDF of the fitted power-law distribution, is depicted in Figures \ref{fig:gamma_dist}-a) and b). It is noticeable that there is a scarcity of high-degree nodes compared to the corresponding power-law distribution and that this pattern is consistent across the three taxonomic domains (as shown in the inset in figures \ref{fig:gamma_dist}-a) and b)). This absence of hubs is a key result of our method as nodes form fewer, but more significant, links in the network.}

\begin{figure}[htbp]
\makebox[\linewidth][c]{
	\centering
\subfloat[Number of Nodes ($N$)]{\includegraphics[width=0.5\textwidth]{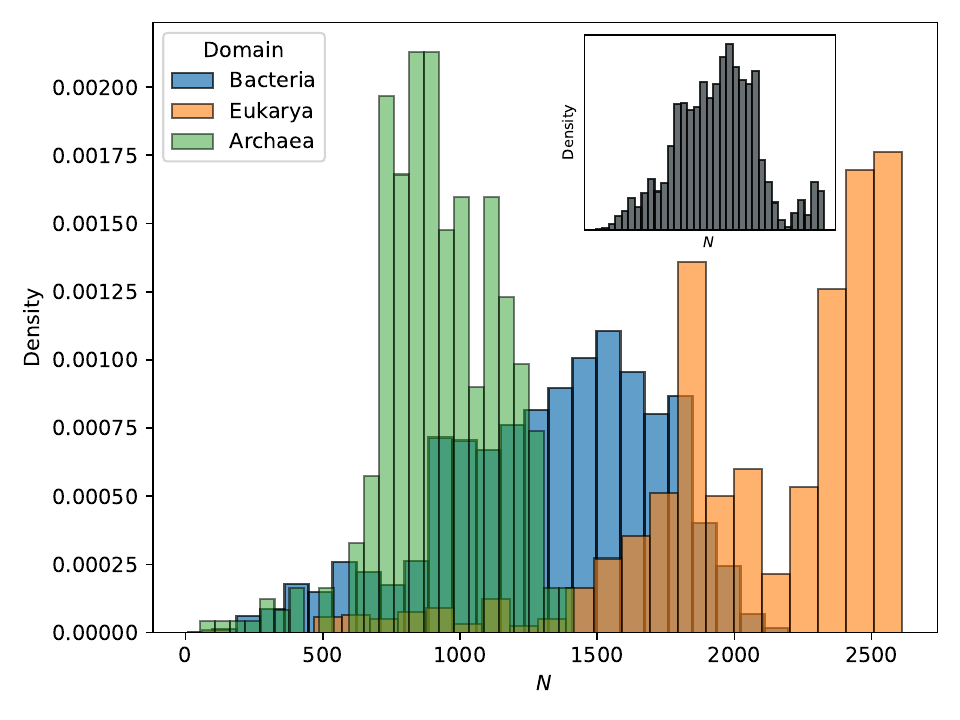}}
\subfloat[Number of Links ($L$)]{\includegraphics[width=0.5\textwidth]{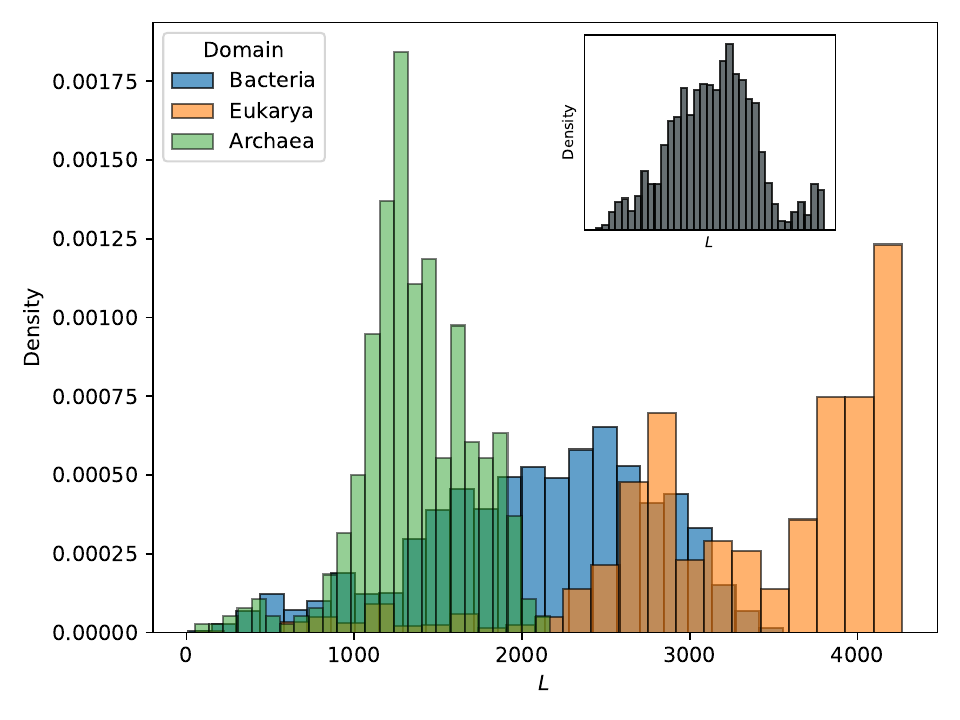}}  }
\makebox[\linewidth][c]{
	\centering
\subfloat[Average Clustering Coefficient ($\bar{C}$)]{\includegraphics[width=0.5\textwidth]{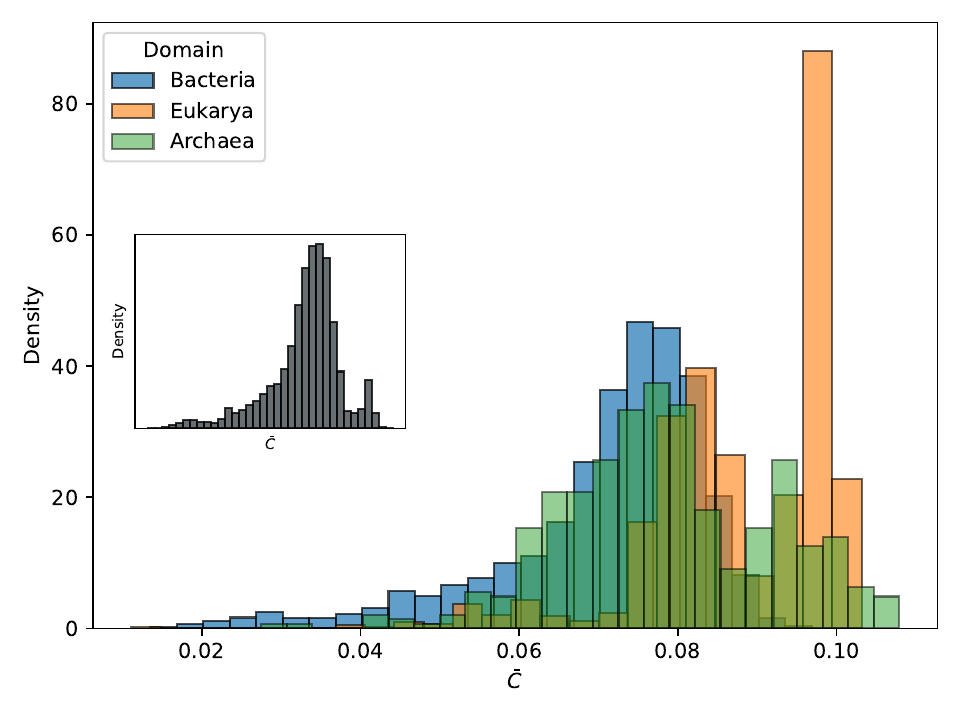}}
\subfloat[Global Clustering Coefficient ($\mathcal{C}$)]{\includegraphics[width=0.5\textwidth]{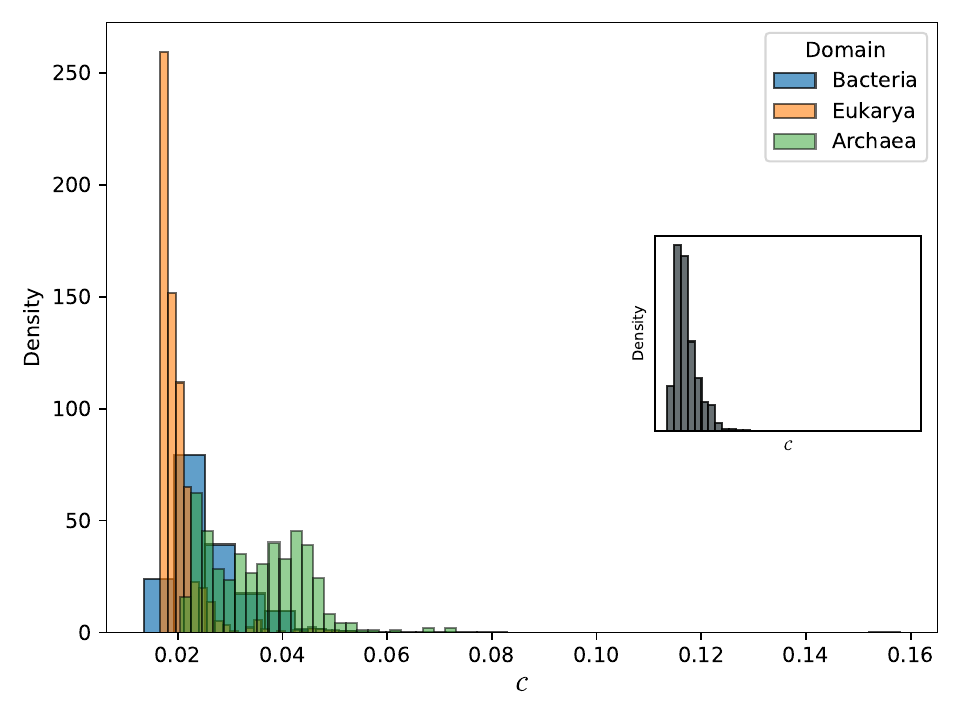}}  }
\makebox[\linewidth][c]{
	\centering
\subfloat[Assortativity ($A$)]{\includegraphics[width=0.5\textwidth]{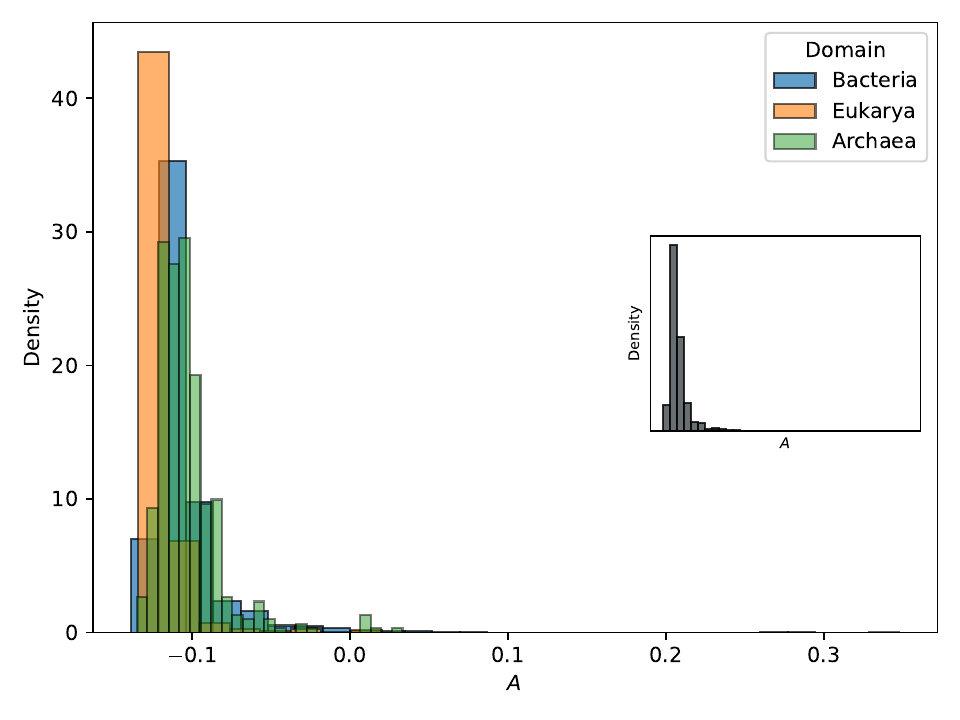}}
\subfloat[Average Distance ($\bar{d}$)]{\includegraphics[width=0.5\textwidth]{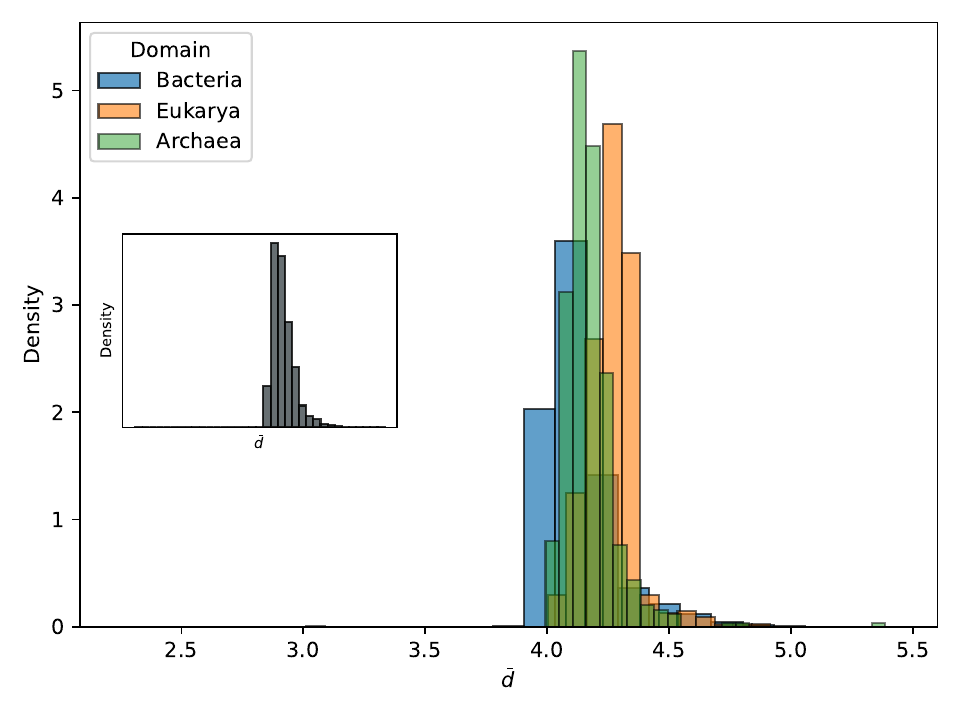}}  }
\caption{{Histograms of parameter distributions in metabolic networks. In gray, we show the parameter distribution of the entire dataset, while in color, we show each Domain separately. For example, in (a), the right-hand peak in the number of nodes distribution (shown in gray) is composed mostly of eukaryotes (in orange).}}
\label{fig:metab_top}
\end{figure}

\begin{figure}[htbp]

\makebox[\linewidth][c]{
    \centering
     
\subfloat[CDF for $k\geq1$.]{\includegraphics[width=0.5\textwidth]{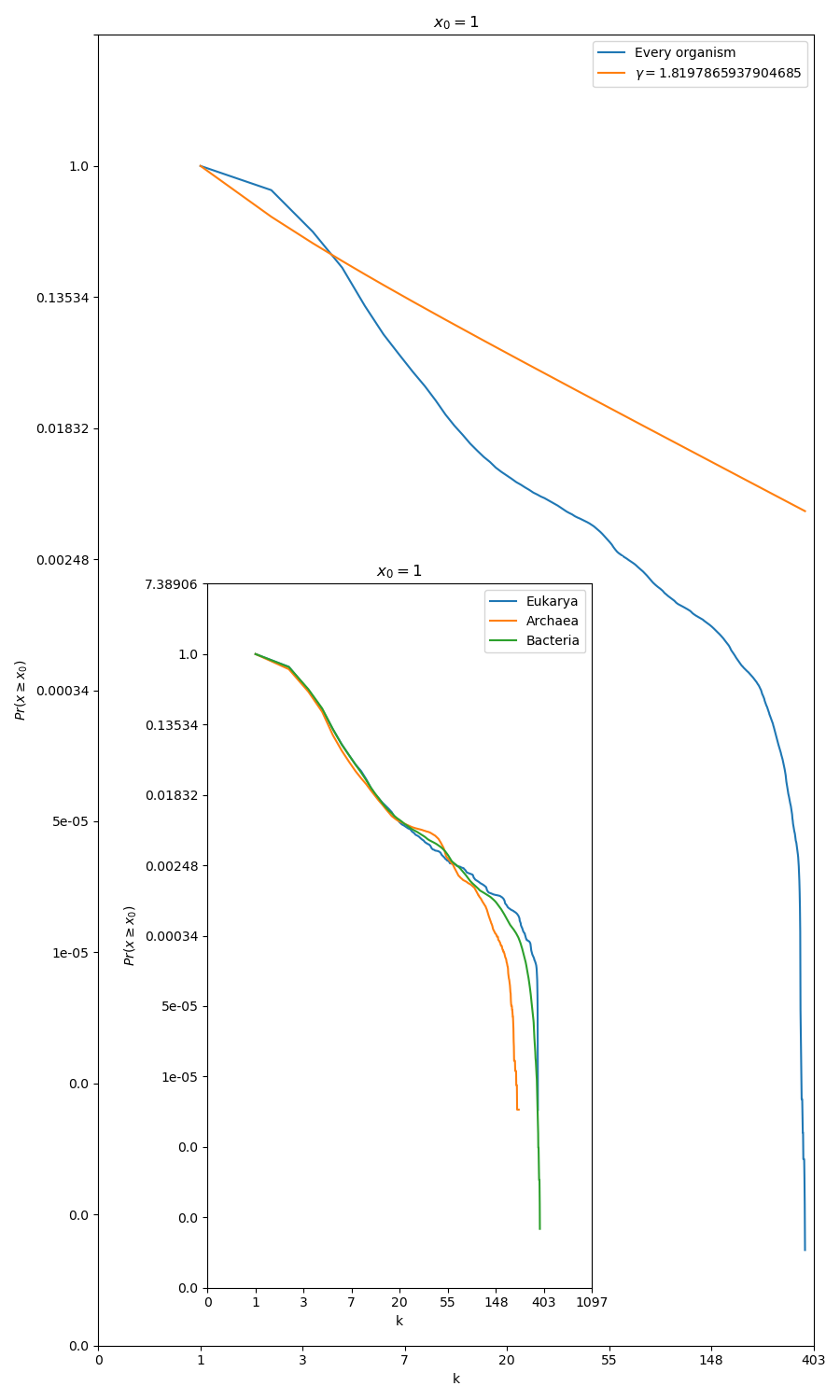}}
     \subfloat[CDF for $k\geq2$.]{\includegraphics[width=0.5\textwidth]{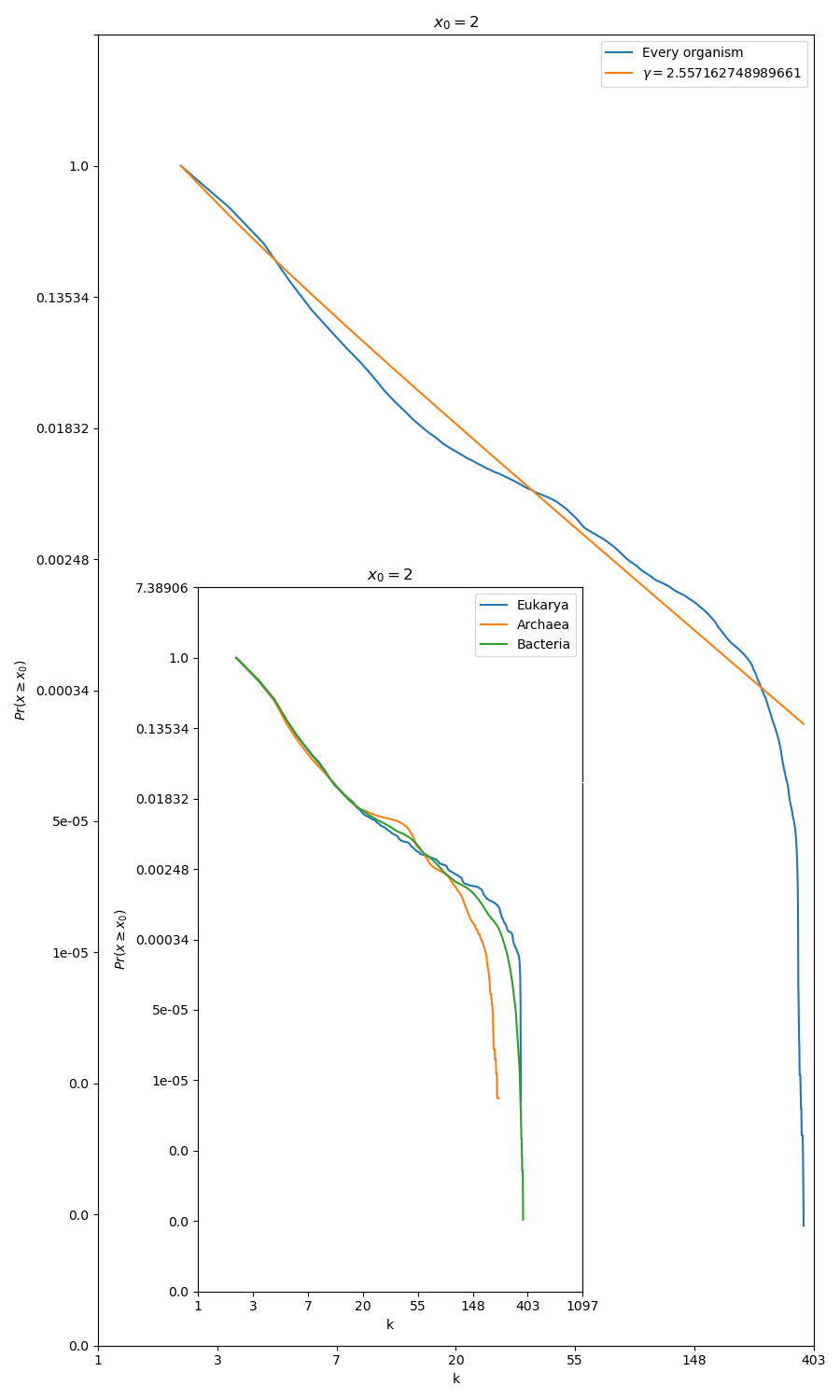}}    } 

\caption{Cumulative distribution function (CDF) for every organism in logarithmic scale. The line represents the CDF of the fitted the power-law distribution for $x_0 = 1$ (a) and $x_0 = 2$ (b). In the insets, we display the CDF for each taxonomic domain. It is evident that all of them exhibit a similar behavior, and there appears to be a scarcity of high-degree nodes compared to the fitted power-law distribution.}
\label{fig:gamma_dist}
\end{figure}

\btab
\centering
\begin{adjustbox}{max width =\textwidth}
\bt{c|ccc}
\hline
$x_0$ & $\gamma$ & Discarded &  Average p-value (fraction $p\geq0.05$) \\ \hline
& & & \\
1 & $1.827914_{-0.024489}^{+0.025064}$ & $0.000000$ & $0.000014 (0.000092)$ \\ 
& & & \\
2 & $2.566756_{-0.055489}^{+0.057068}$ & $0.315579$ & $0.000769 (0.003391)$ \\ 
& & & \\
3 & $2.881485_{-0.090488}^{+0.094063}$ & $0.637136$ & $0.000197 (0.000733)$ \\ 
& & & \\
4 & $3.039050_{-0.127424}^{+0.133841}$ & $0.791657$ & $0.000128 (0.000458)$ \\ 
& & & \\
5 & $2.825116_{-0.150996}^{+0.161191}$ & $0.884564$ & $0.000192 (0.001100)$ \\ 
& & & \\
6 & $2.681764_{-0.170869}^{+0.185258}$ & $0.925045$ & $0.000173 (0.000550)$ \\ 
& & & \\
7 & $2.621357_{-0.190862}^{+0.209937}$ & $0.945061$ & $0.000157 (0.000367)$ \\ 
& & & \\
8 & $2.561103_{-0.207609}^{+0.230660}$ & $0.957840$ & $0.000134 (0.000183)$ \\ 
& & & \\
9 & $2.516804_{-0.223565}^{+0.252010}$ & $0.966216$ & $0.000168 (0.000550)$ \\ 
& & & \\
10 & $2.449374_{-0.236062}^{+0.268815}$ & $0.972659$ & $0.000196 (0.000733)$ \\ 
& & & \\
11 & $2.350198_{-0.242960}^{+0.281012}$ & $0.977885$ & $0.000151 (0.000642)$ \\ 
& & & \\
12 & $2.298004_{-0.253570}^{+0.296639}$ & $0.981211$ & $0.000177 (0.001009)$ \\ 
& & & \\
13 & $2.255551_{-0.261436}^{+0.310003}$ & $0.983679$ & $0.000194 (0.001192)$ \\ 
& & & \\
14 & $2.215618_{-0.269110}^{+0.323373}$ & $0.985682$ & $0.000211 (0.000825)$ \\ 
& & & \\
15 & $2.196400_{-0.278360}^{+0.338214}$ & $0.987110$ & $0.000239 (0.001284)$ \\ 
& & & \\
\hline
    \et
    \end{adjustbox}
    \caption{Results of the fitting procedure for the power-law distribution. Here $x_0$ represents the minimum degree considered in the fitting process. The corresponding fraction of discarded nodes is shown in the third column.}
    \label{tab:gamma}
    \etab

Now, data on community structure parameters are available in Table \ref{tab:metab_surp}. Additionally, histograms for these parameters are presented in Figure \ref{fig:metab_com}. We note that the community structure of metabolic networks is made up of many small communities, since the average community size is small $\bar{n}_c = 3.64 \pm 0.20$ and the average number of communities is large $N_c = 381.15 \pm 118.16$. Furthermore, we see that the communities are fairly evenly-sized as the average Pielou's index is close to one $\overline{PI} = 0.961 \pm 0.004$. Since Surprise is not affected by the resolution limit known to bias modularity-based community detection methods, the relatively large number of communities observed here may reflect the presence of genuine small-scale structures that remained unresolved in previous studies \cite{Fortunato36}. An important point is that the communities identified in this work should not be interpreted as equivalent to complete metabolic pathways. Community detection is based exclusively on network topology, whereas metabolic pathways are functional annotations that may span several interconnected modules. The relatively small community sizes observed here should, therefore, be interpreted in the context of the network representation adopted in this work. By linking metabolites only through chemically meaningful transformations, our approach removes numerous indirect connections generated by reaction co-participation and currency metabolites, resulting in a substantially less interconnected graph. Consequently, community detection identifies modules that are smaller and more localized than those typically observed in traditional substrate networks. Furthermore, because the Surprise quality function does not suffer from the resolution limit associated with Modularity, these communities may represent small-scale structures that remained unresolved in previous studies. Rather than corresponding to entire biochemical pathways, the detected communities are more likely to represent compact transformation modules that are connected to one another through a limited number of inter-community links. The relatively small and homogeneous community sizes therefore suggest that metabolism is organized around many localized transformation units that are subsequently integrated into larger functional processes. Such an organization is consistent with the hierarchical modularity previously reported for metabolic networks, where small chemically coherent modules combine to form larger functional structures and pathways \cite{ravasz2002hierarchical}.

\btab[htbp]
    \centering
    \begin{adjustbox}{max width=0.8\textwidth}
    \bt{p{4cm}c|ccc}
    \hline
    Parameter & Math symbol & Average & Standard deviation & Skewness \\
    \hline
     & & & & \\
    Number of possible links inside communities & $M$ & $3584.144062_{-1428.715897}^{+1689.745837}$ & $1561.175208$ & $0.590557$ \\ 
  & & & & \\ 
Actual number of links inside communities & $\zeta$ & $1216.332478_{-430.313566}^{+435.909357}$ & $433.153110$ & $0.207848$ \\ 
  & & & & \\ 
Number of communities & $N_c$ & $381.154051_{-124.245204}^{+112.443417}$ & $118.158107$ & $-0.118160$ \\ 
  & & & & \\ 
Surprise & $S$ & $5634.436011_{-2141.769250}^{+2294.280413}$ & $2219.928000$ & $0.359201$ \\ 
  & & & & \\ 
Average community size & $\bar{n}_c$ & $3.639186_{-0.225858}^{+0.168959}$ & $0.196728$ & $-0.811420$ \\ 
  & & & & \\ 
Size of biggest community & $\max(n_c)$ & $32.422654_{-8.871701}^{+6.751427}$ & $7.794539$ & $-0.643897$ \\ 
  & & & & \\ 
Pielou's Index & ${PI}$ & $0.961726_{-0.004010}^{+0.004776}$ & $0.004394$ & $0.585475$ \\ 
  & & & & \\ 
Number of Bridges & $N_{B}$ & $12.805260_{-4.928308}^{+5.455128}$ & $5.196762$ & $0.357757$ \\ 
  & & & & \\ 
\hline
    \et
    \end{adjustbox} 
 \caption{Characteristics of the community structure within metabolic networks.} 
\label{tab:metab_surp} 
\etab

\begin{figure}

\makebox[\linewidth][c]{
    \centering
     
\subfloat[Surprise ($S$).]{\includegraphics[width=0.5\textwidth]{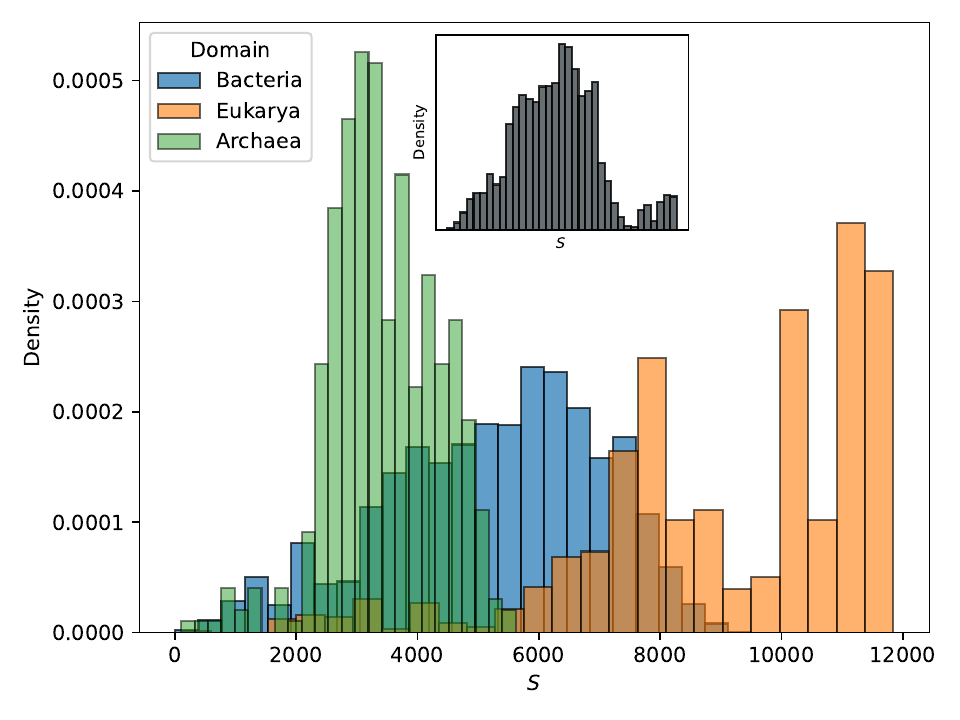}}
     \subfloat[Pielou's Index ($PI$).]{\includegraphics[width=0.5\textwidth]{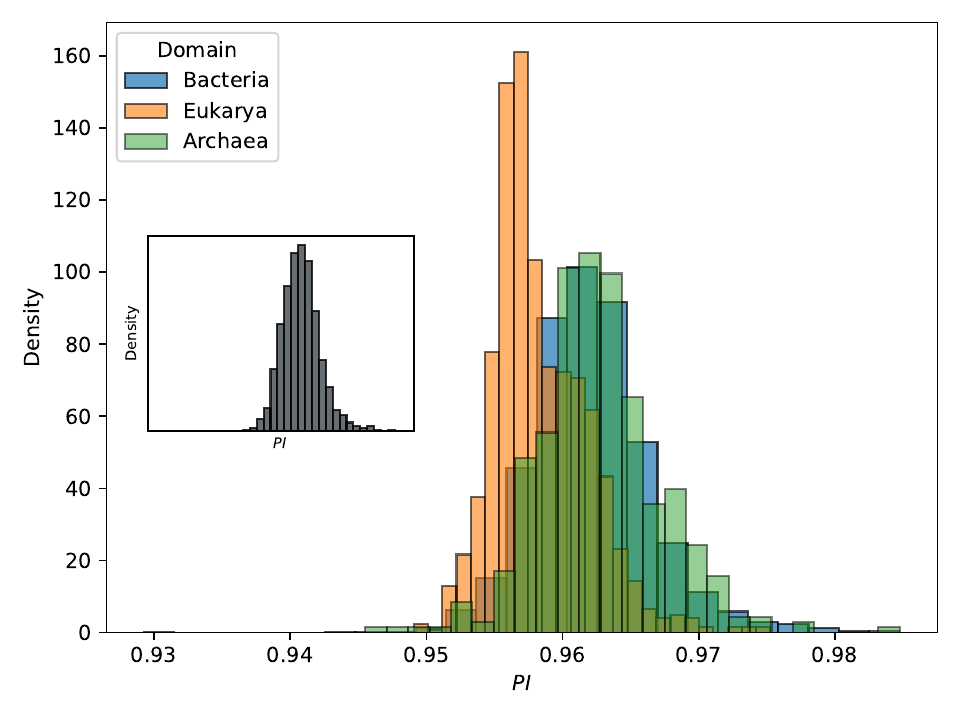}}    } 
\makebox[\linewidth][c]{
    \centering
     
\subfloat[Number of Communities ($N_c$).]{\includegraphics[width=0.5\textwidth]{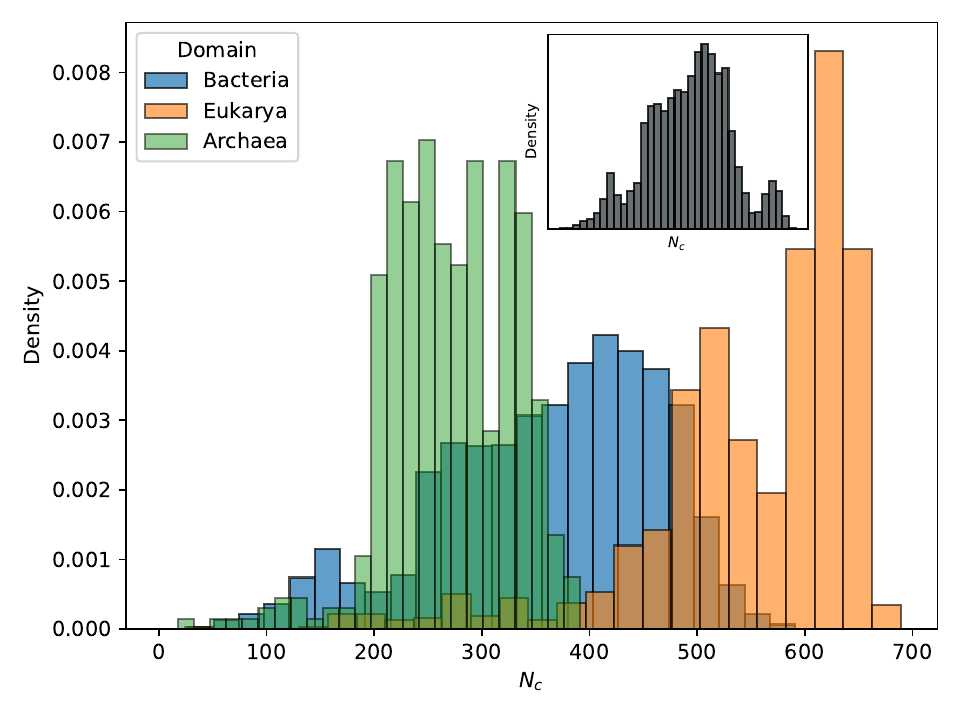}}
     \subfloat[Average Community Size ($\bar{n}_c$).]{\includegraphics[width=0.5\textwidth]{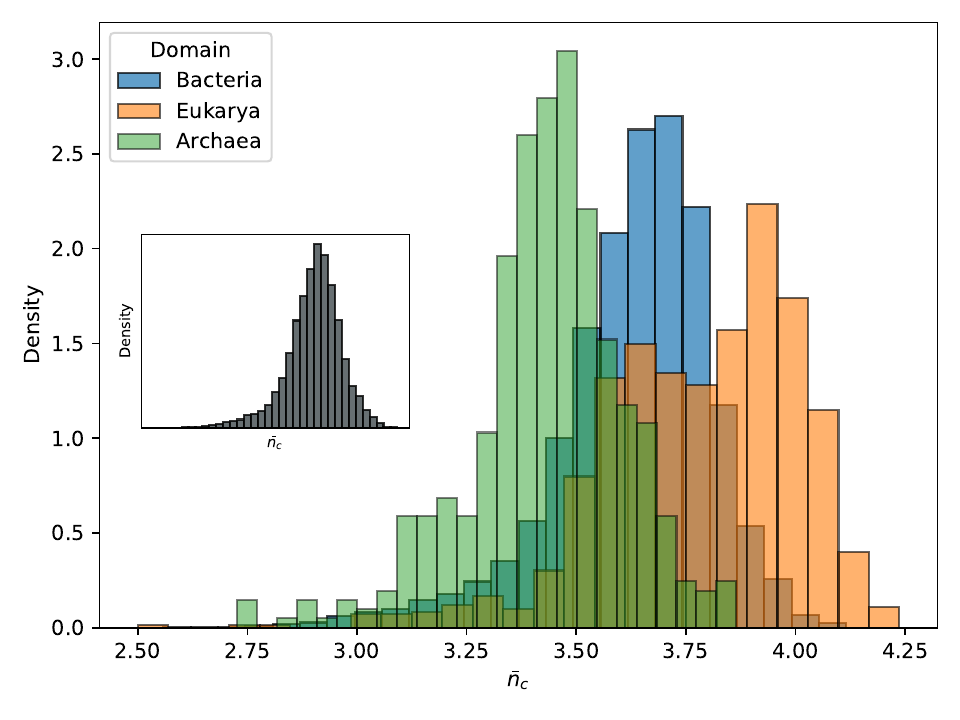}}    } 

\makebox[\linewidth][c]{
    \centering
     
\subfloat[Maximum Community Size $\max(n_c)$.]{\includegraphics[width=0.5\textwidth]{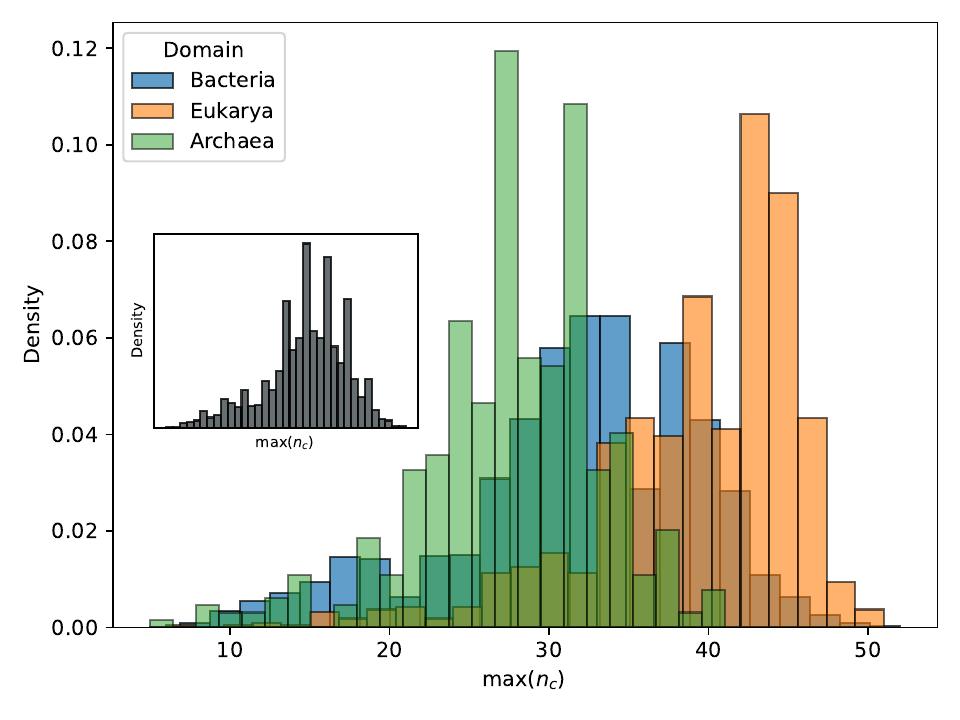}}
     \subfloat[Proportion of intra-community links $\frac{\zeta}{L}$.]{\includegraphics[width=0.5\textwidth]{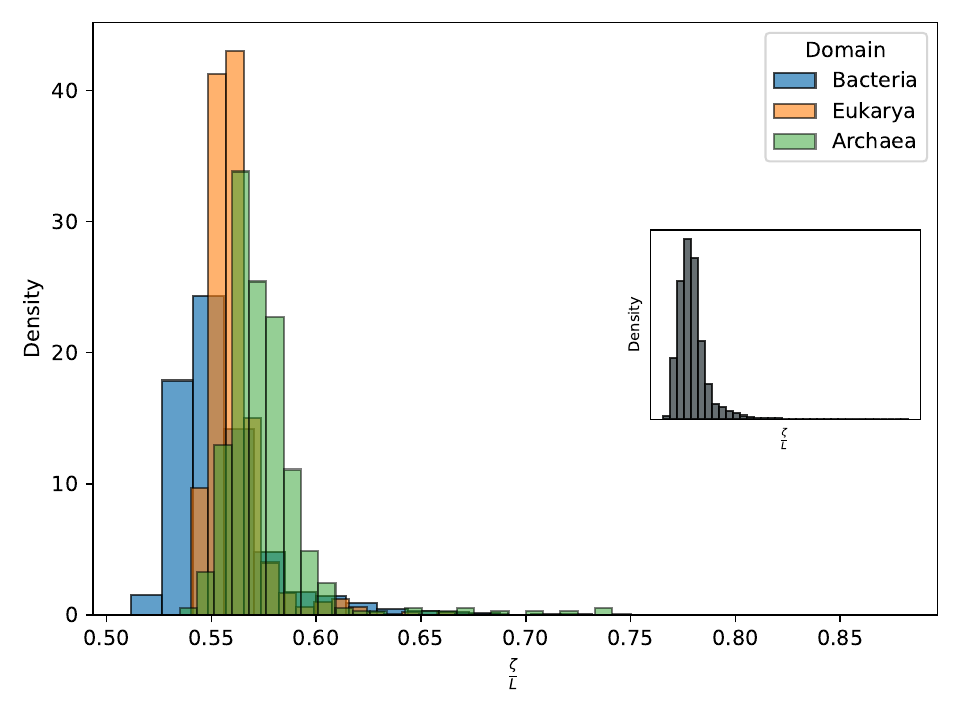}}    } 

\caption{{Histograms of the community structure distributions in metabolic networks. In gray, we show the parameter distribution of the entire dataset, while in color, we show each Domain separately. For instance, in d), we see a very even distribution in gray-scale, but, when we filter by domain we can see a separation between Archaea, Bacteria and Eukaryotes.}}
    \label{fig:metab_com}
\end{figure}

\subsection{Comparison between real and randomized networks}

\btab[htbp]
    \centering
    \begin{adjustbox}{max width=0.8\textwidth}
    \bt{p{4cm}c|ccc}
    \hline
    Parameter & Math symbol & Average & Standard deviation & Skewness \\
    \hline
     & & & & \\
    Average local clustering & $\bar{C}_{\textrm{rand}}$ & $0.038226_{-0.008755}^{+0.005996}$ & $0.007300$ & $-0.472415$ \\ 
  & & & & \\ 
Global Clustering & $\mathcal{C}_{\textrm{rand}}$ & $0.014457_{-0.002180}^{+0.004523}$ & $0.003359$ & $21.797294$ \\ 
  & & & & \\ 
Assortativity & $A_{\textrm{rand}}$ & $-0.113614_{-0.006822}^{+0.013651}$ & $0.009964$ & $1.883361$ \\ 
  & & & & \\ 
Average distance & $\bar{d}_{\textrm{rand}}$ & $4.334003_{-0.114229}^{+0.272651}$ & $0.186802$ & $2.402436$ \\ 
  & & & & \\ 
Number of components & $N_{\textrm{comps rand}}$ & $26.600110_{-5.820602}^{+7.252096}$ & $6.521653$ & $0.502749$ \\ 
  & & & & \\ 
Size of main component & ${\mathcal{S}}_{\textrm{main rand}}$ & $1341.783413_{-469.342917}^{+455.319791}$ & $462.085055$ & $0.078617$ \\ 
  & & & & \\ 
Average size of smaller components & $\mathcal{S}_{\textrm{small rand}}$ & $2.269901_{-0.061239}^{+0.083307}$ & $0.072391$ & $1.731503$ \\ 
  & & & & \\ 
\hline
    \et
    \end{adjustbox} 
 \caption{Topological properties of randomized networks.} 
\label{tab:rand_top} 
\etab

\btab[htbp]
    \centering
    \begin{adjustbox}{max width=0.8\textwidth}
    \bt{p{4cm}c|ccc}
    \hline
    Parameter & Math symbol & Average & Standard deviation & Skewness \\
    \hline
     & & & & \\
    Number of possible links inside communities & $M$ & $3624.536364_{-1365.752820}^{+1392.200690}$ & $1379.233688$ & $0.224753$ \\ 
  & & & & \\ 
Actual number of links inside communities & $\zeta$ & $1109.323011_{-382.958871}^{+372.492765}$ & $377.532691$ & $0.093586$ \\ 
  & & & & \\ 
Number of communities & $N_c$ & $368.779032_{-125.549877}^{+120.777342}$ & $123.095378$ & $0.018734$ \\ 
  & & & & \\ 
Surprise & $S$ & $4949.961506_{-1868.027192}^{+1967.463830}$ & $1918.933657$ & $0.304327$ \\ 
  & & & & \\ 
Average community size & $\bar{n}_c$ & $3.798601_{-0.133260}^{+0.117301}$ & $0.125330$ & $-0.576060$ \\ 
  & & & & \\ 
Size of biggest community & $\max(n_c)$ & $34.133761_{-8.620032}^{+5.890034}$ & $7.202179$ & $-0.893654$ \\ 
  & & & & \\ 
Pielou's Index & $PI$ & $0.955859_{-0.001938}^{+0.004202}$ & $0.002978$ & $2.060627$ \\ 
  & & & & \\ 
Number of Bridges & $N_{B}$ & $28.156965_{-8.816225}^{+7.924727}$ & $8.356810$ & $-0.140752$ \\ 
  & & & & \\ 
\hline
    \et
    \end{adjustbox} 
 \caption{Characteristics of the community structure of randomized networks.} 
\label{tab:rand_surp} 
\etab

The statistical characteristics of the randomized networks are presented in Tables \ref{tab:rand_top} and \ref{tab:rand_surp}, while histograms illustrating the distribution of $t_p$ values for the topological properties are depicted in Figures \ref{fig:t_percent_top_1} and \ref{fig:t_percent_top_2}. The dashed lines in the plots of $t_p$ represent the critical values of $t_p$. Networks that fall between these dashed lines have a p-value higher than 0.05 for the characteristic in question. For networks outside these dashed lines, a negative value of $t_p$ indicates that the parameter value in the real metabolic network is higher than the expected value from the randomized networks. Conversely, a positive value of $t_p$ tells us that the parameter value of the real network is smaller than the expected value from the randomized networks.

Supplementary to the histograms of $t_p$, we computed, for each domain, the percentage of networks exhibiting p-values exceeding 0.05 for the respective parameter. These percentages are depicted in the bar plots situated in the rightmost column of the figures.

\begin{figure}[htbp]

\makebox[\linewidth][c]{
    \centering
     
\subfloat[Average Local Clustering Coefficient ($t_{\bar{C}}$).]{\includegraphics[width=0.59\textwidth]{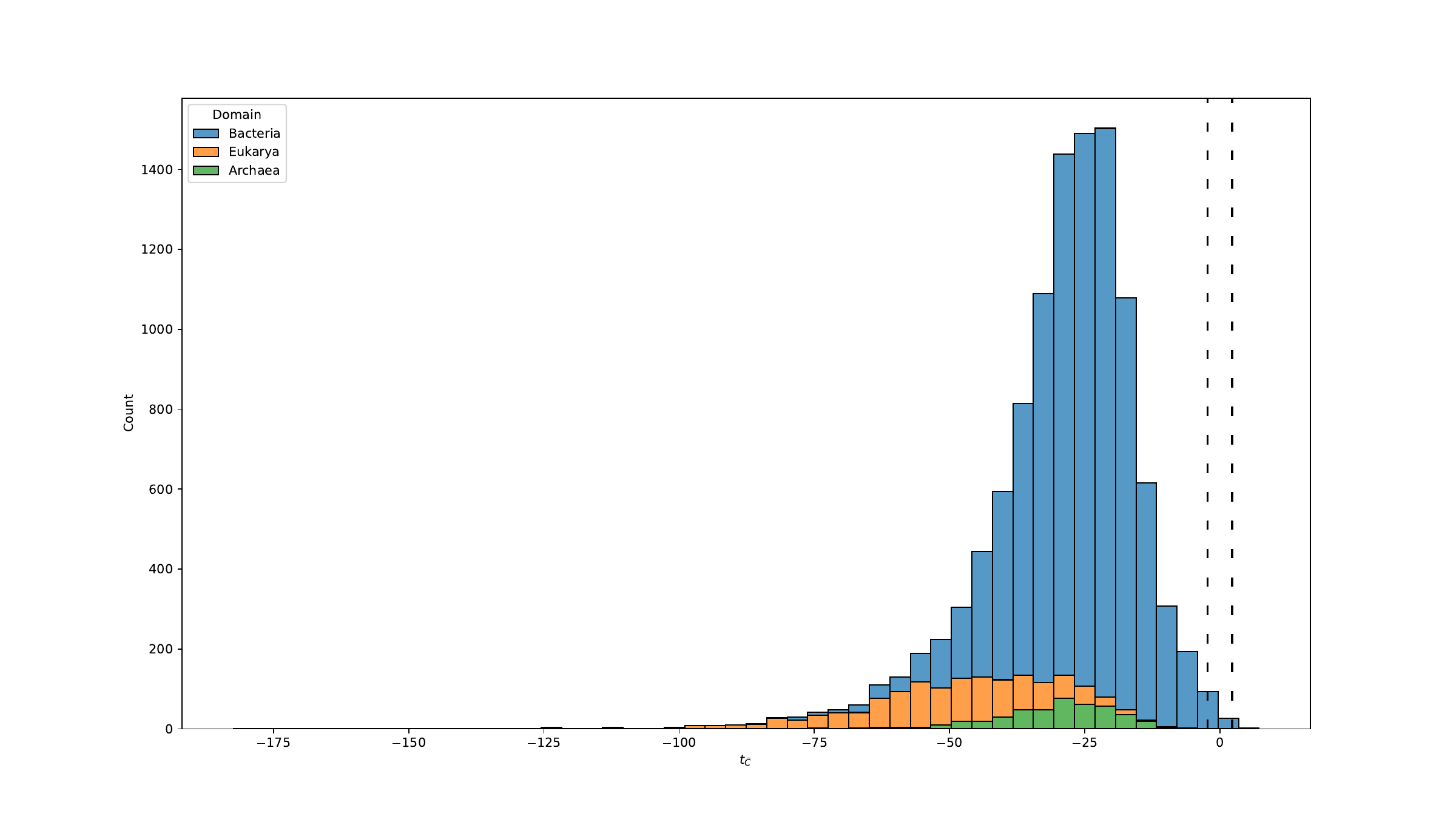}}
     \subfloat[Fraction p$<0.05$ per domain.]{\includegraphics[width=0.41\textwidth]{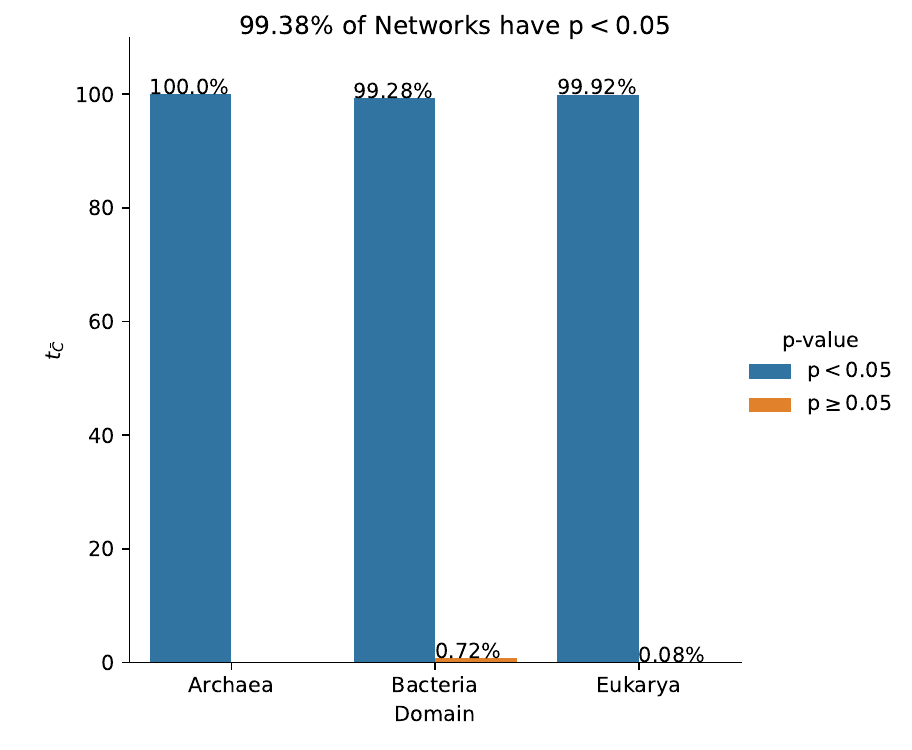}}    } 
\makebox[\linewidth][c]{
    \centering
     
\subfloat[Global Clustering Coefficient ($t_{\mathcal{C}}$).]{\includegraphics[width=0.59\textwidth]{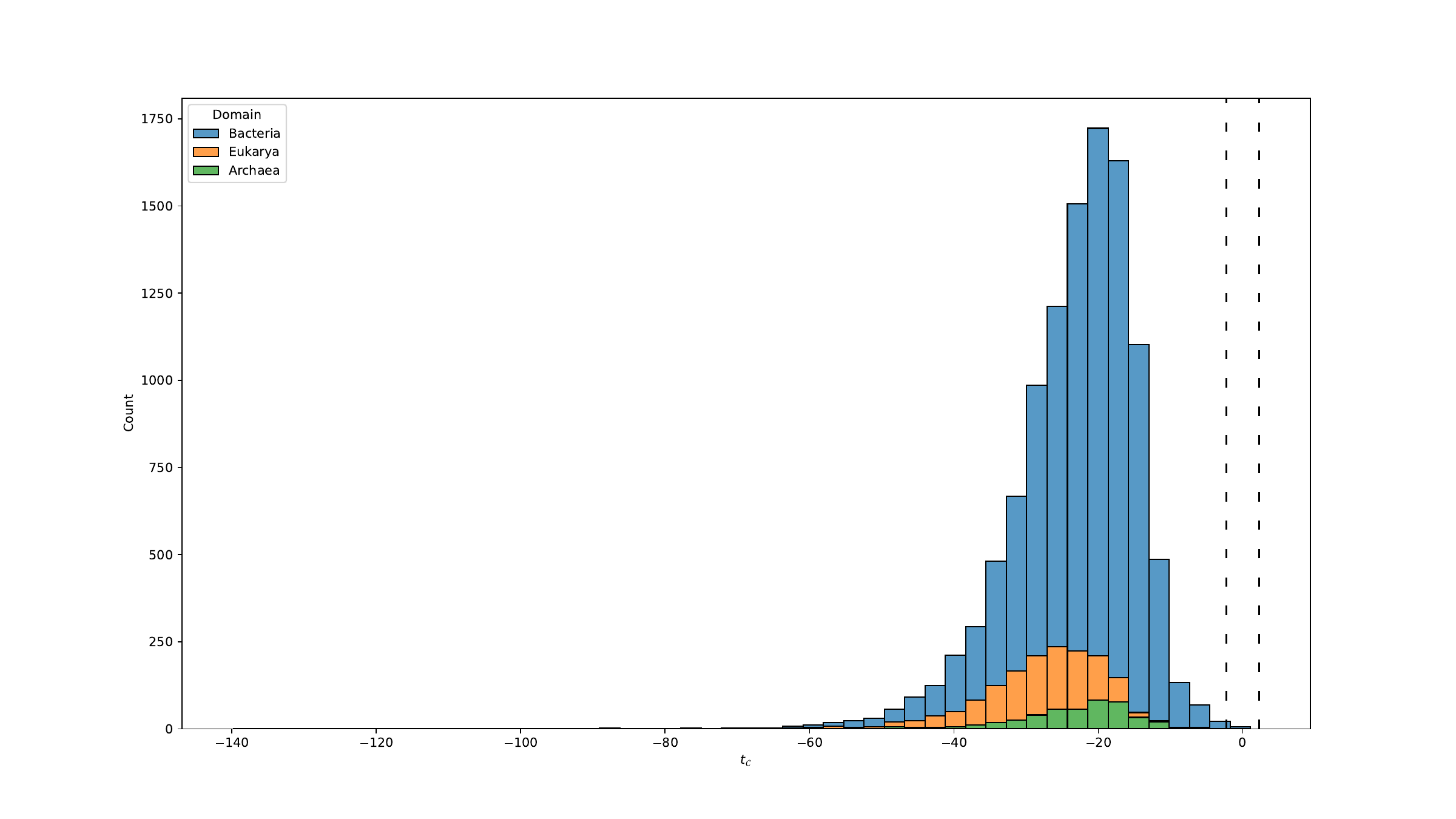}}
     \subfloat[Fraction p$<0.05$ per domain.]{\includegraphics[width=0.41\textwidth]{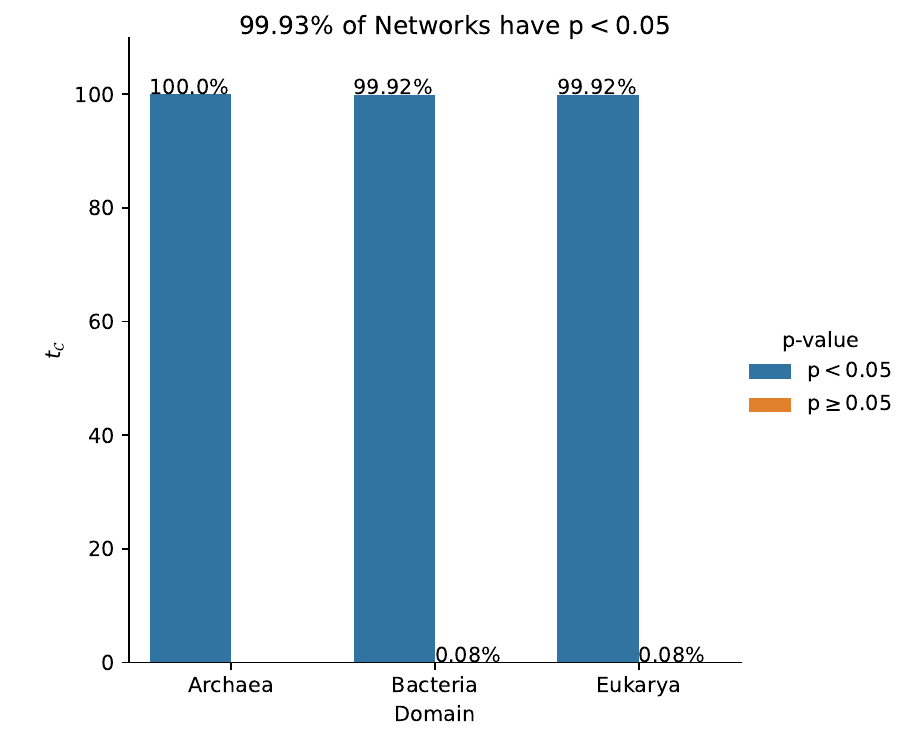}}    } 

\makebox[\linewidth][c]{
    \centering
     
\subfloat[Average Distance ($t_{\bar{d}}$).]{\includegraphics[width=0.59\textwidth]{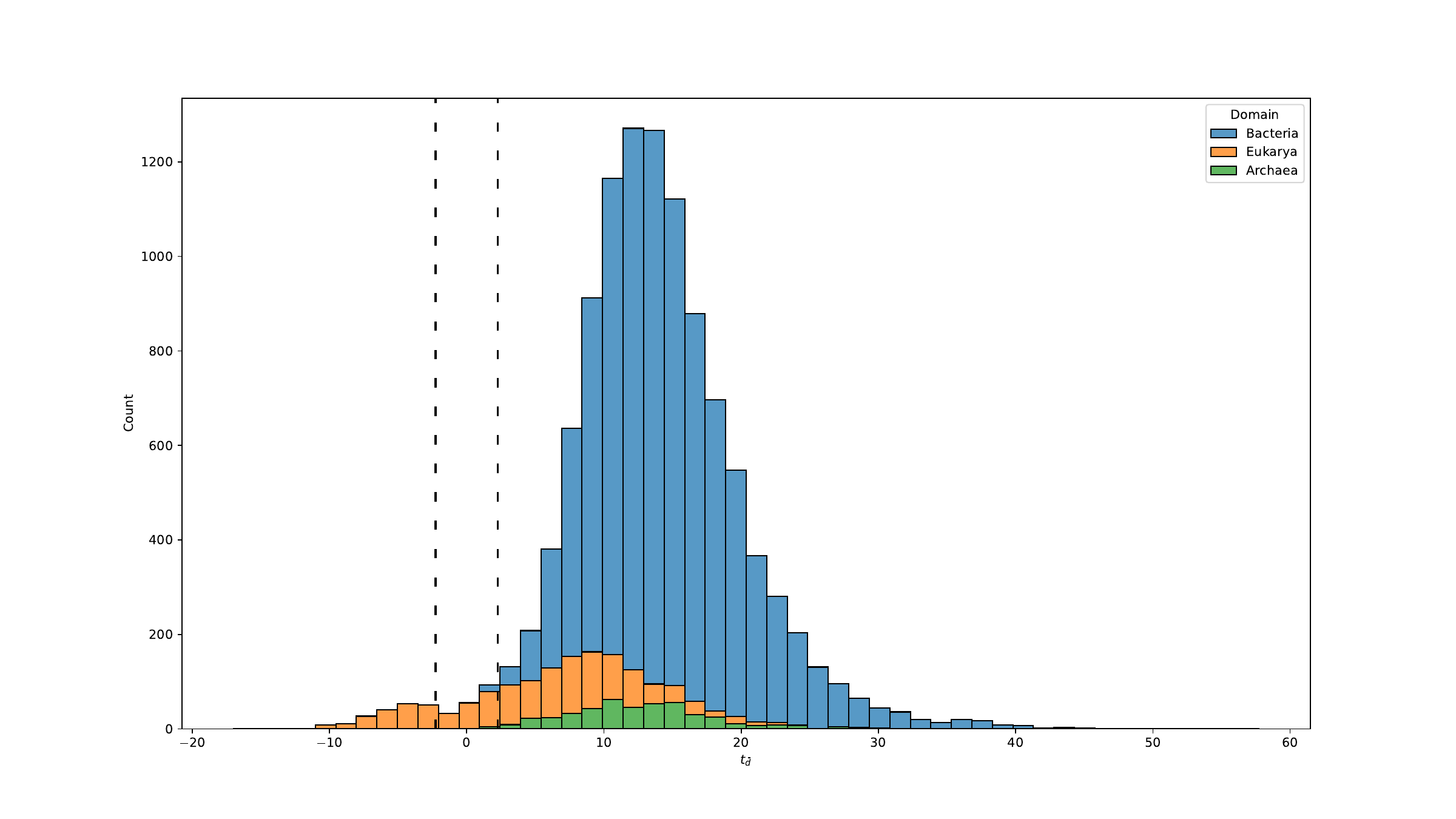}}
     \subfloat[Fraction p$<0.05$ per domain.]{\includegraphics[width=0.41\textwidth]{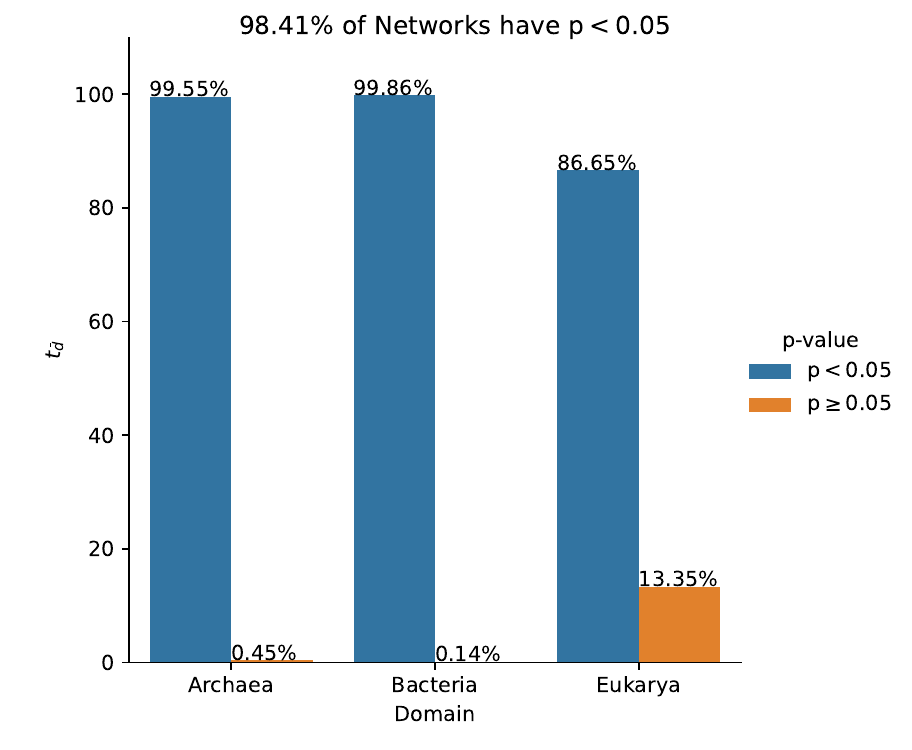}}    }

\caption{Histograms of $t_p$ for the topological properties of the metabolic networks. Here, we choose to stack the distributions for better visibility. In the right side we show the percentage of networks with p$<0.05$ per domain.}
    \label{fig:t_percent_top_1}
\end{figure}

\begin{figure}[htbp]

\makebox[\linewidth][c]{
    \centering
     
\subfloat[Assortativity ($t_{A}$).]{\includegraphics[width=0.59\textwidth]{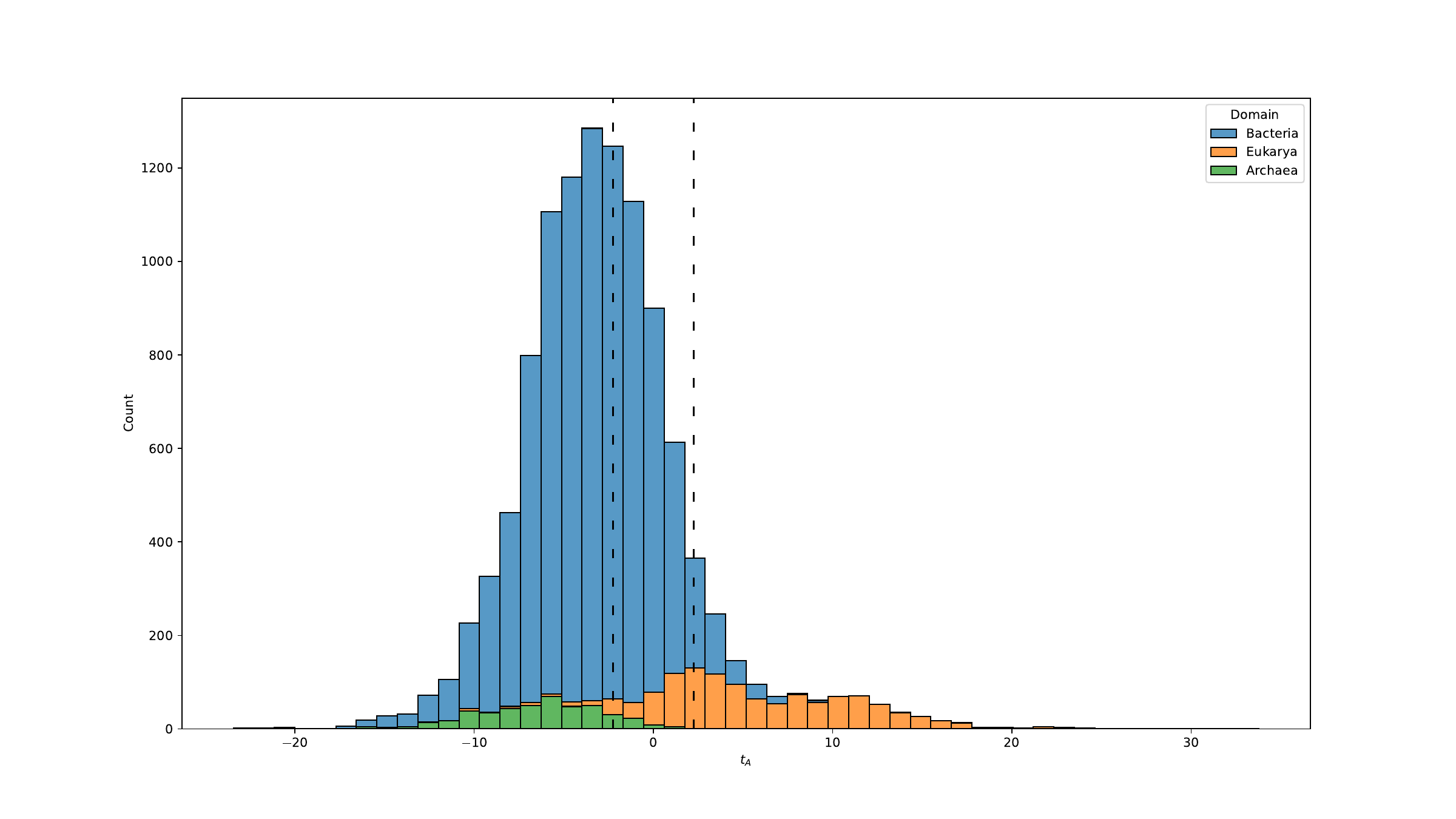}}
     \subfloat[Fraction p$<0.05$ per domain.]{\includegraphics[width=0.41\textwidth]{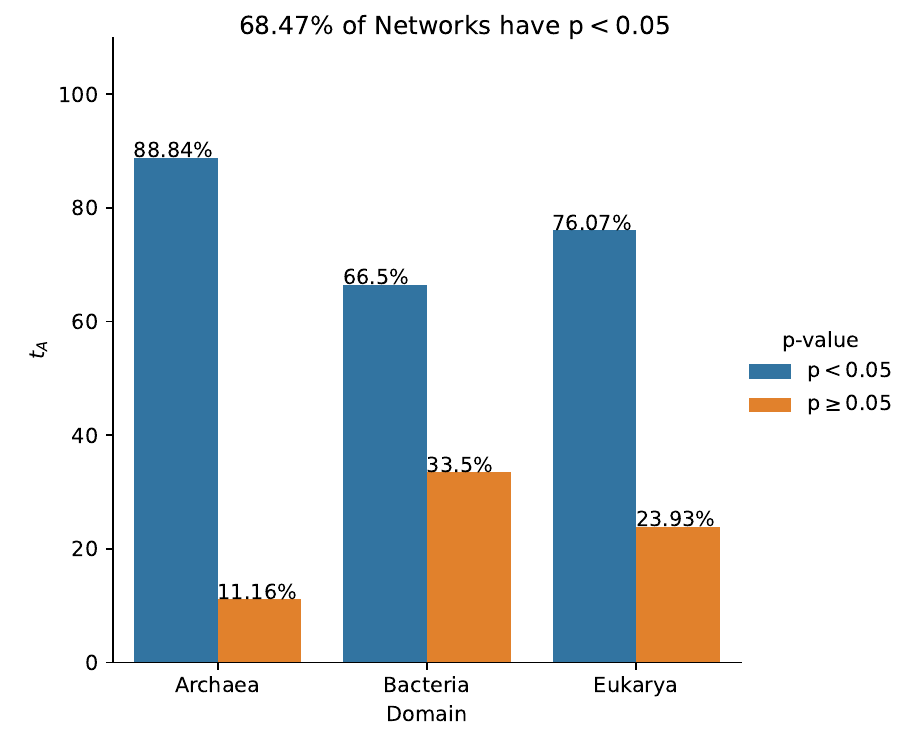}}    } 
 
\caption{Histograms of $t_p$ for the topological properties of the metabolic networks. Here, we choose to stack the distributions for better visibility. On the right side, we show the percentage of networks with p$<0.05$ per domain.}
    \label{fig:t_percent_top_2}
\end{figure}

In relation to the network's topological properties, we see that the most ubiquitous differences are in the local and global clustering coefficients $\bar{C}$ and $\mathcal{C}$, respectively (figures \ref{fig:t_percent_top_1}-a to -d). The deviation between the real values and the expected values from the randomized versions is statistically significant for more than $99.3\%$ of the networks within the dataset, employing a confidence threshold of $95\%$.

Furthermore, we observe that almost all distributions exhibit a negative value for $t_p$, suggesting that actual metabolic networks possess higher local and global clustering coefficients compared to their randomized counterparts. Remember that the statistical significance does not provide information about the magnitude of the difference between real and randomized parameters; it only indicates the confidence level at which we can state that a difference exists.

We can, however, grasp the magnitude of this difference by comparing the average values of the real parameters with the expected values from the randomized samples. With respect to the local clustering coefficient, we note that the value of $\bar{C}$ in randomized networks is approximately half ($52.21\%$) the value in real metabolic networks. This does indicate that real networks show a tendency to cluster locally and form tightly connected groups: the communities we are interested in.

Although it is true that the global clustering coefficient $\mathcal{C}$ is also higher in real networks, we must remember that the value of $\mathcal{C}$ is much smaller than the value of $\bar{C}$ and thus its influence is smaller. Furthermore, the difference between $\mathcal{\bar{C}_{\textrm{rand}}}$ and $\bar{C}_{\textrm{rand}}$ is proportionally smaller than the difference between $\mathcal{C}$ and $\bar{C}$. This also suggests that the observed clusters are the result of different mechanisms and not a mere result of the degree distribution of the network.

{Regarding the remaining parameters, only $1.59\%$ of organisms do not display a statistically significant difference in average distance. Among the rest, $96.67\%$ exhibit a lower-than-expected average distance $\bar{d}$ compared to their randomized counterparts, while only $1.73\%$ show a higher-than-expected value. However, this latter group is composed exclusively of organisms from the Eukarya domain, and, in particular, of organisms from the Animalia kingdom. Indeed, the Animalia kingdom displays a distinct pattern compared to all other kingdoms in the dataset: only $54.27\%$ of the animals have a lower average distance than expected, $20.53\%$ show no statistically significant differences, and $25.2\%$ exhibit a higher-than-expected $\bar{d}$. This could indicate that more complex organisms have a less compact metabolic network than expected from their degree distribution.} {Numerous factors influence the characteristics of metabolic networks in complex organisms, including specialization, regulatory mechanisms, and evolutionary history. Complex organisms possess specialized metabolic pathways and compartments, resulting in a network that is less interconnected overall. Moreover, the presence of sophisticated regulatory mechanisms introduces additional layers of control, giving rise to a more distributed pattern of connections within the network \cite{klein2020emergence}. Additionally, the distinct evolutionary pathways taken by an organism can play a crucial role in shaping the architecture of its metabolic network \cite{takemoto2012current}}.

{The differences in Assortativity are not as ubiquitous, with only $68.45\%$ of organisms showing a statistically significant value of $t_A$. As with the average distance, however, clear distinctions emerge between the three domains. The metabolic networks of bacteria and archaea tend to be less disassortative than their randomized counterparts, indicating that assortativity $A$ is generally higher in real networks.
This trend reverses in the Eukarya domain: $70.78\%$ of organisms from the Eukarya domain exhibit \( t_A \geq 0 \) (indicating that assortativity \( A \) is smaller in real networks), while only \( 5.29\% \) show \( t_A \leq 0 \). The remaining \( 23.93\% \) do not display a statistically significant result difference.
Within Eukarya, the trend becomes even clearer: $82.66\%$ of animals, 
$75.58\%$ of plants, and 
$45.79\%$ of fungi exhibit lower-than-expected assortativity, compared to only 
$1.43\%$ of protists. These results highlight a structural divergence between the metabolic organization of multicellular eukaryotes and that of prokaryotes, suggesting that increasing {the} organism complexity may be associated with a reduction in assortativity.}

\begin{figure}[htbp]

\makebox[\linewidth][c]{
    \centering
     
\subfloat[Surprise ($t_{S}$).]{\includegraphics[width=0.59\textwidth]{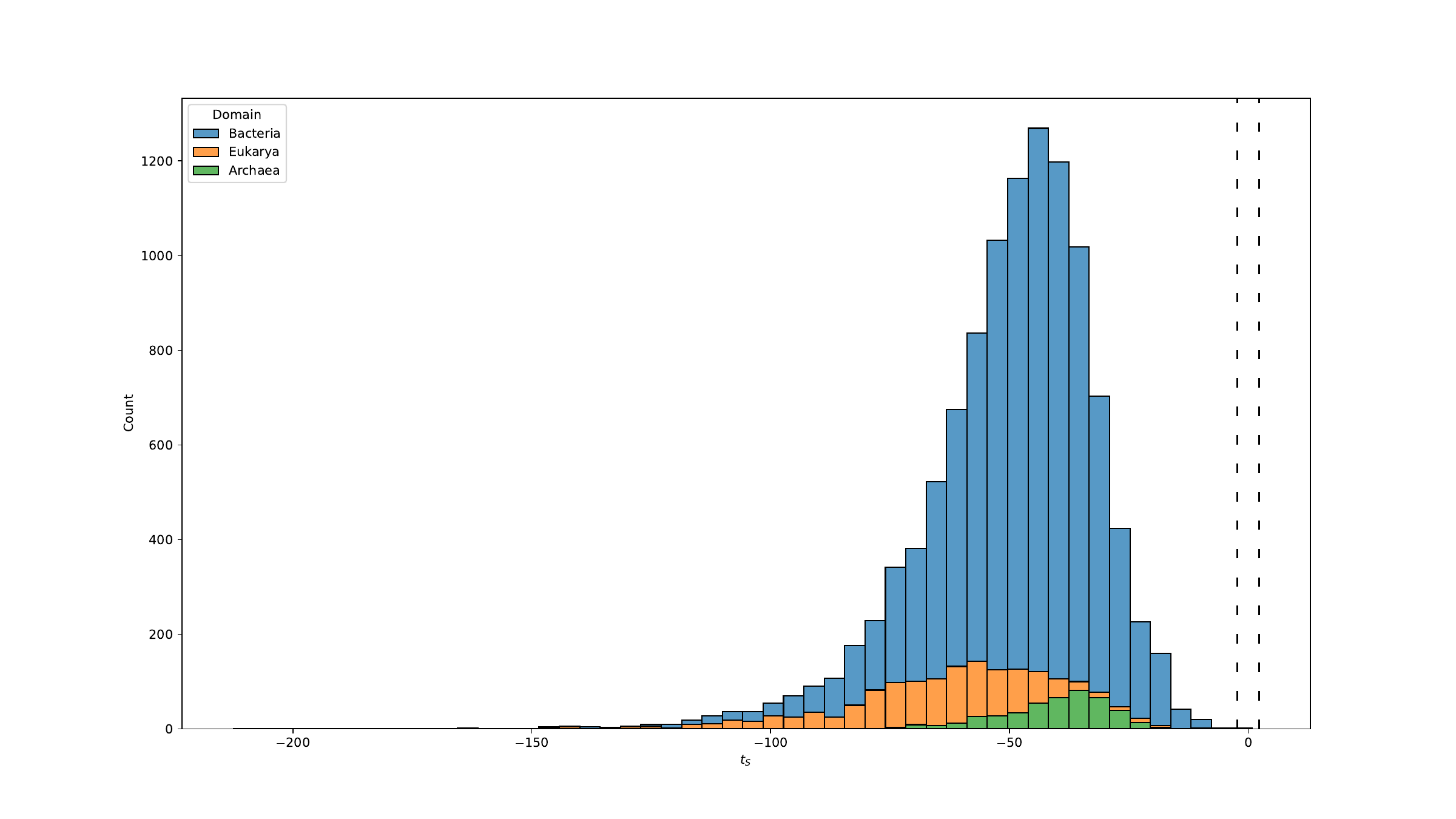}}
     \subfloat[Fraction of $p<0.05$ per domain.]{\includegraphics[width=0.41\textwidth]{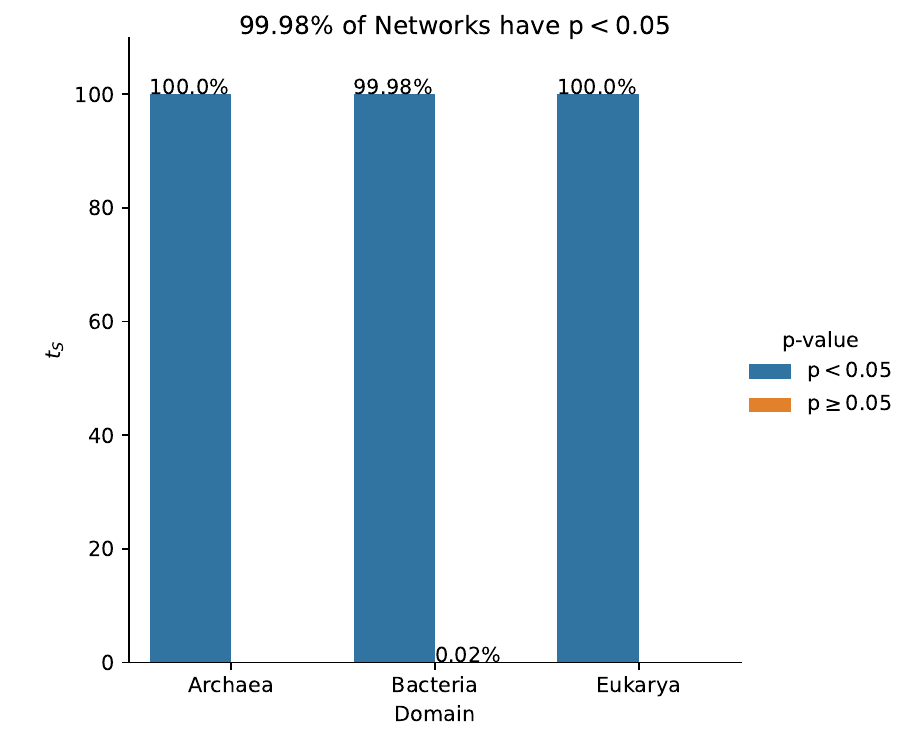}}    } 
\makebox[\linewidth][c]{
    \centering
     
\subfloat[Number of possible intra-community links ($t_{M}$).]{\includegraphics[width=0.59\textwidth]{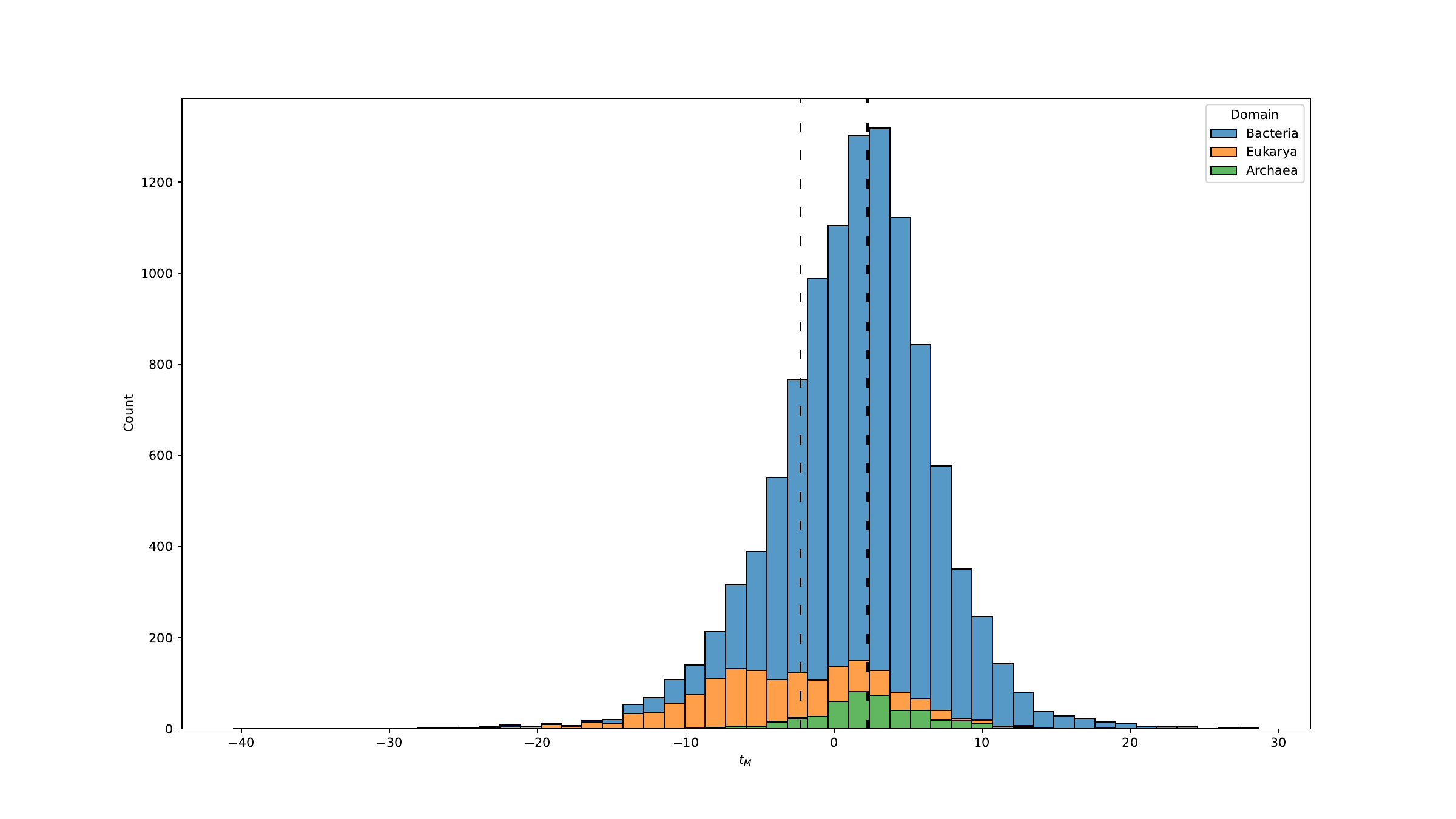}}
     \subfloat[Fraction of $p<0.05$ per domain.]{\includegraphics[width=0.41\textwidth]{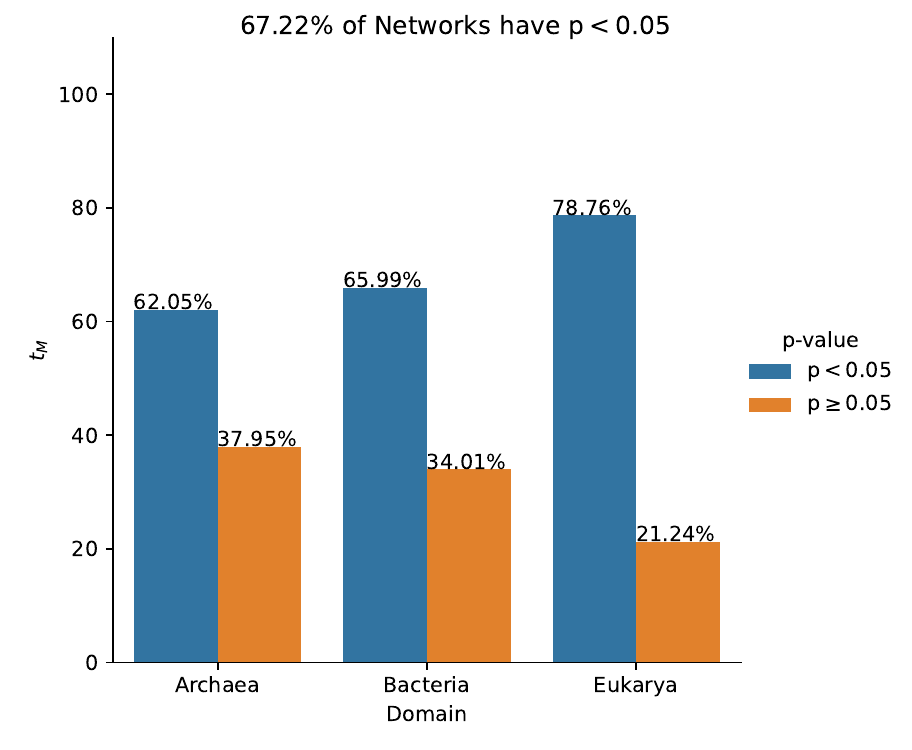}}    } 

\makebox[\linewidth][c]{
    \centering
     
\subfloat[Number of actual intra-community links ($t_{\zeta}$).]{\includegraphics[width=0.59\textwidth]{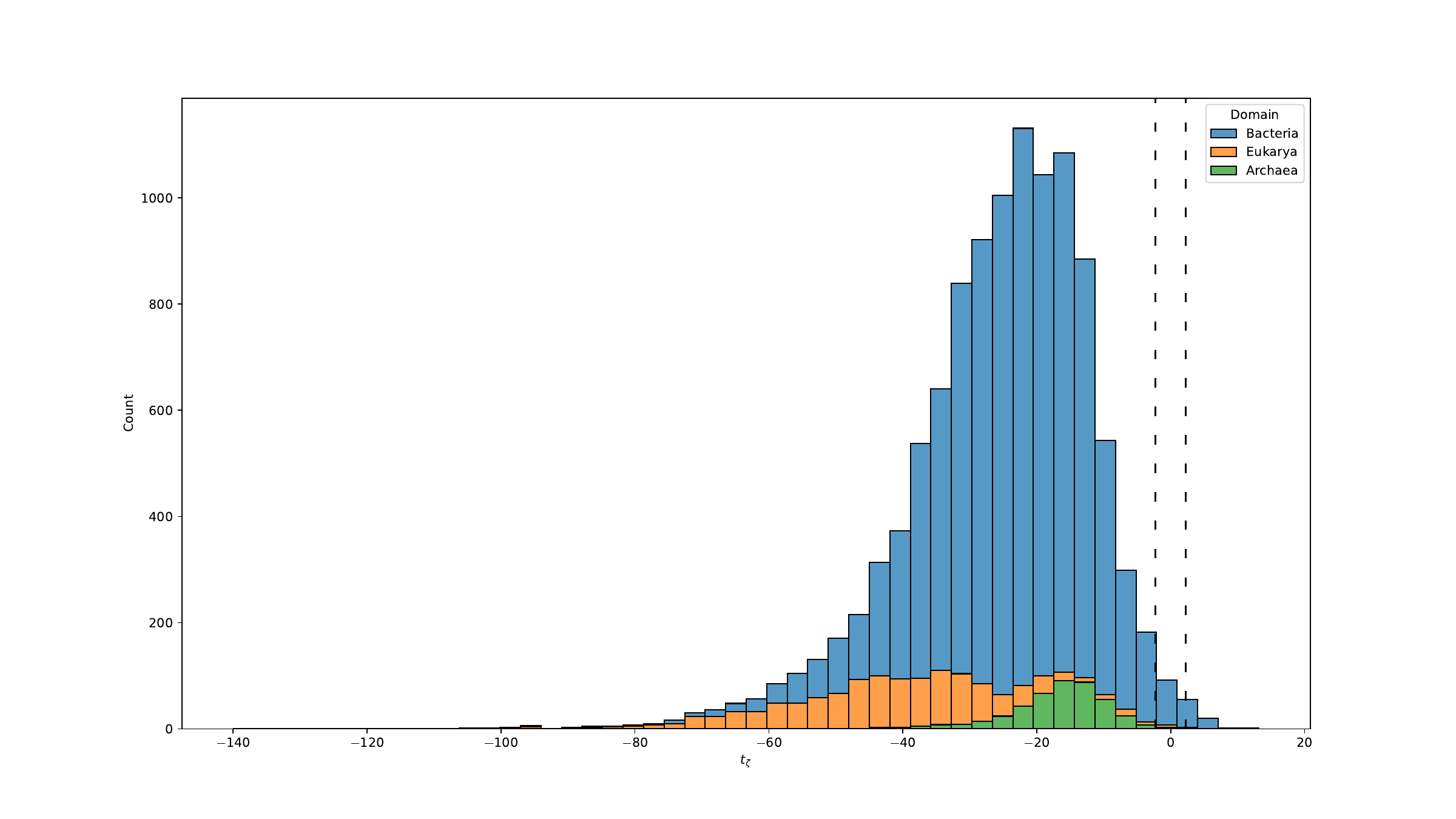}}
     \subfloat[Fraction of $p<0.05$ per domain.]{\includegraphics[width=0.41\textwidth]{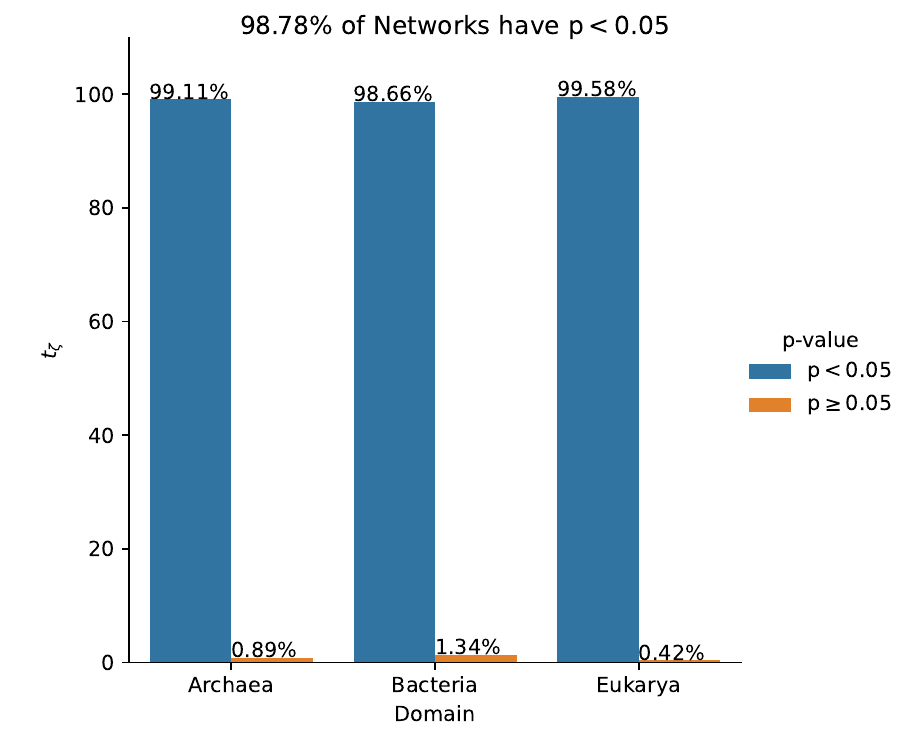}}    } 

\caption{Histograms displaying $t_p$ for the community structure parameters of the metabolic networks. Once again, we choose to stack the distributions for better visibility. On the right side, we show the percentage of networks with p$<0.05$ per domain.}
    \label{fig:t_percent_surp}
\end{figure}

In relation to the community structure, figures \ref{fig:t_percent_surp}-a to -f show the histograms of $t_p$ for the parameters associated to the Surprise function. The initial observation is that virtually all metabolic networks exhibit a higher Surprise value compared to the randomized networks. {This result, no doubt, correlates with the higher local clustering coefficient in the real networks. Higher local clustering means more tightly knit small groups of nodes densely connected.} {In fact, of the 10912 organisms studied, only 2 show a p-value that is not statistically significant. This implies that real-world networks possess a stronger community structure, since the Surprise function quantifies the level of surprise (or unlikeliness) in obtaining, at random, a partition with the same enrichment of intra-community links. When examining the 2 organisms that do not meet this condition, we find that they belong to two distinct phyla: \textit{Bacillota} (\textit{Firmicutes}) and \textit{Flavobacteria}. The former, represented by \textit{Carboxydichorda subterranea}\cite{karnachuk2024novel}, is a thermophilic bacterium isolated from a deep terrestrial aquifer in western Siberia, whereas the latter, represented by \textit{Bostrichicola ureolyticus}\cite{kiefer2023cuticle}, is a co-obligate symbiont of beetles from the family Bostrichidae, contributing to cuticle biosynthesis in its host. As both possess some of the smallest metabolic networks in the dataset, we can say that the absence of a statistically significant difference in the Surprise between their community structures and those of their randomized counterparts can likely be attributed to the simplicity of their metabolic networks, arising from adaptations to extreme environmental conditions in \textit{C. subterranea} and from metabolic dependency on the host in \textit{B. ureolyticus}.}

Proceeding with our analysis, we see that the number of intra-community links $\zeta$ found in the metabolic networks is higher than their randomized versions for over $98.78\%$ of organisms. This tells us that nodes are more likely to have connections inside their community compared to randomized versions of the network. If we combine this finding with the fact that the average local clustering and the values of Surprise are higher in the real networks, it is clear that the metabolic networks do present a community structure and that the community structure of those networks is not solely the result of the degree distribution of the network but of other evolutionary pressures as well.

\begin{figure}[htbp]

\makebox[\linewidth][c]{
    \centering
     
\subfloat[Number of communities ($t_{N_c}$).]{\includegraphics[width=0.59\textwidth]{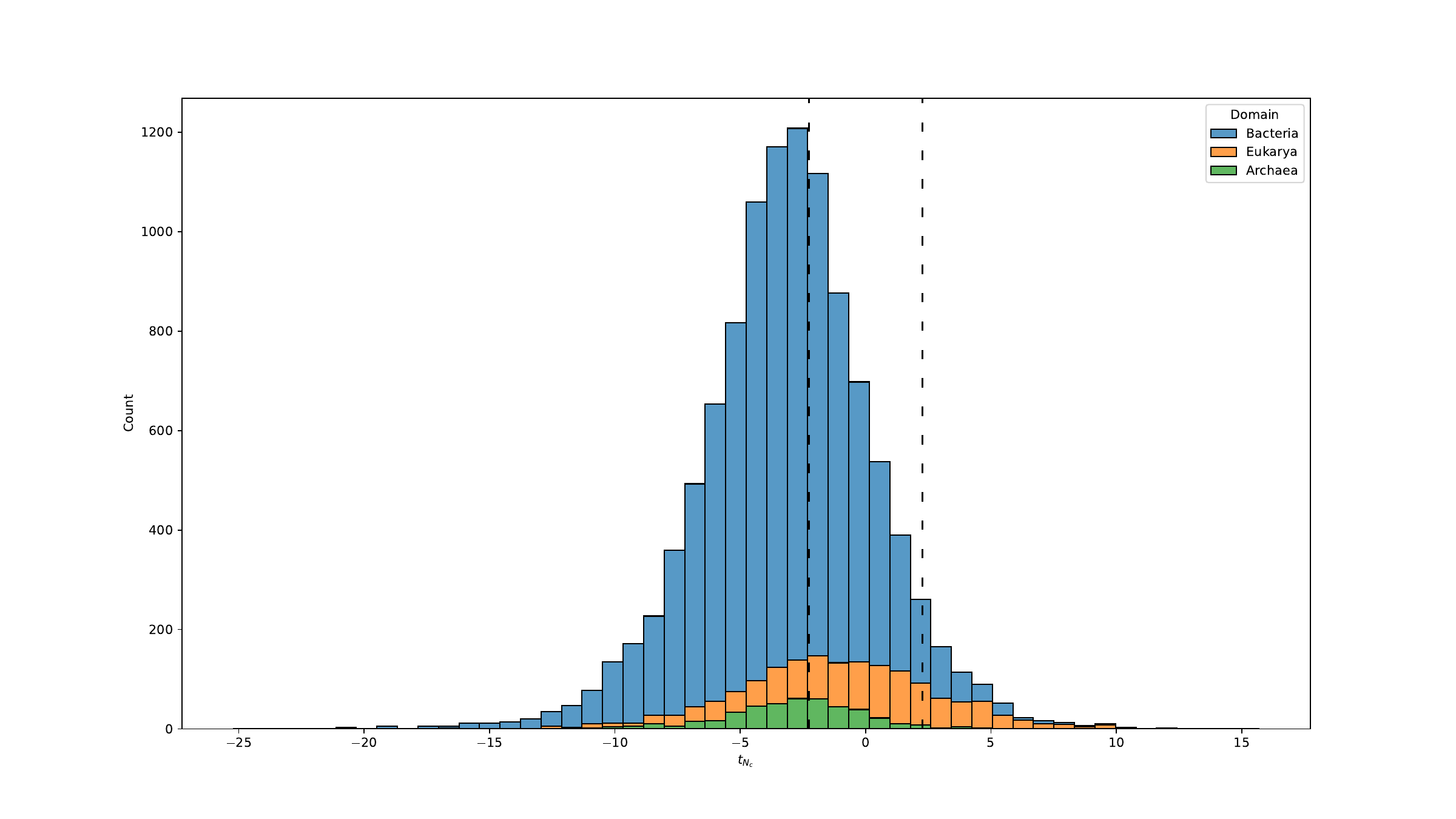}}
     \subfloat[Fraction of $p<0.05$ per domain.]{\includegraphics[width=0.41\textwidth]{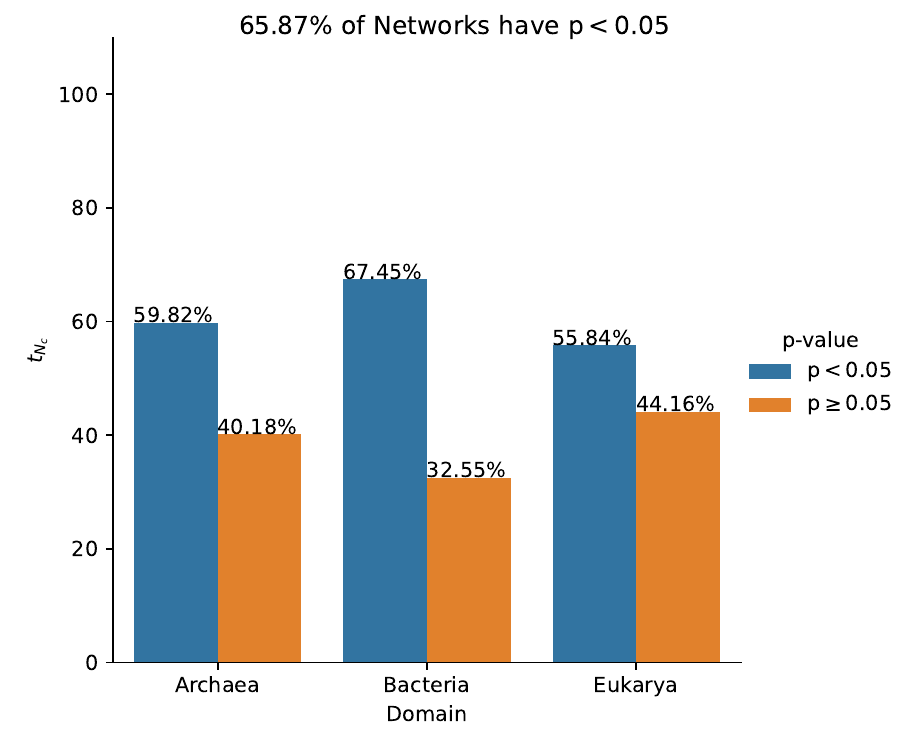}}    } 
\makebox[\linewidth][c]{
    \centering
     
\subfloat[Average Community Size ($t_{\bar{n}_c}$).]{\includegraphics[width=0.59\textwidth]{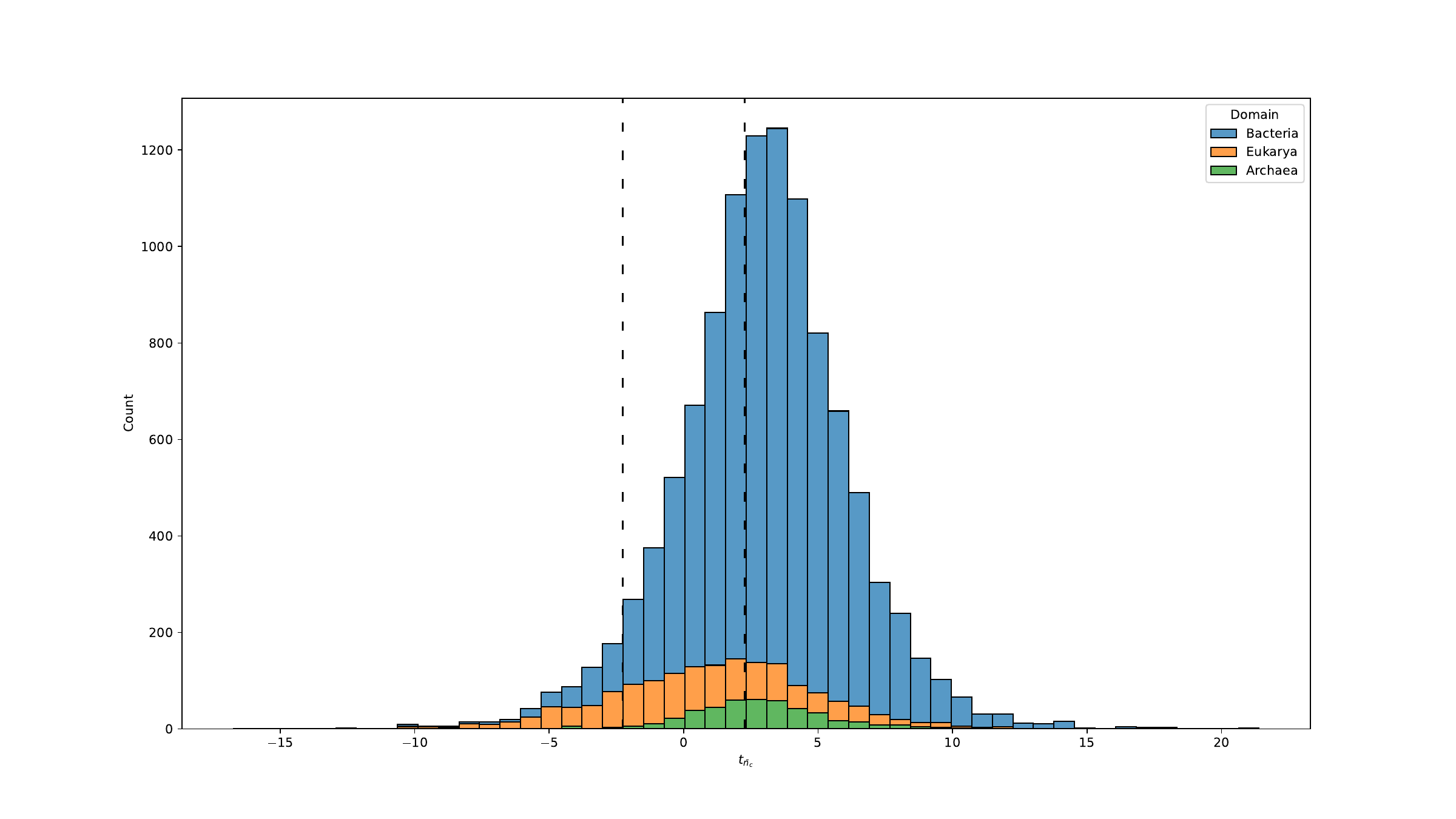}}
     \subfloat[Fraction of $p<0.05$ per domain.]{\includegraphics[width=0.41\textwidth]{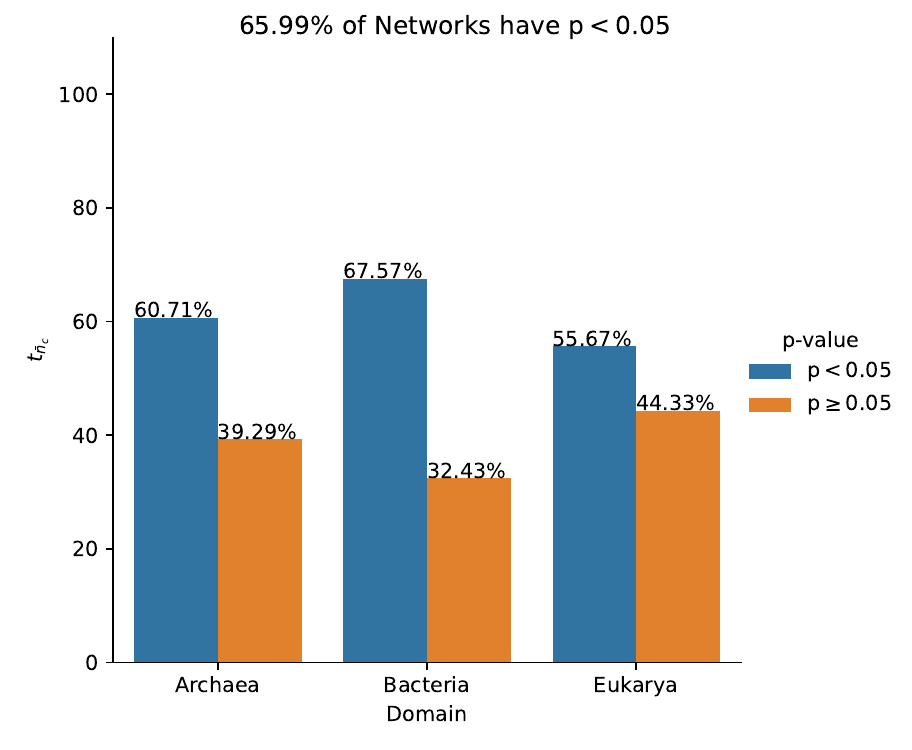}}    } 

\makebox[\linewidth][c]{
    \centering
     
\subfloat[Maximum Community Size ($t_{\max(n_c)}$).]{\includegraphics[width=0.59\textwidth]{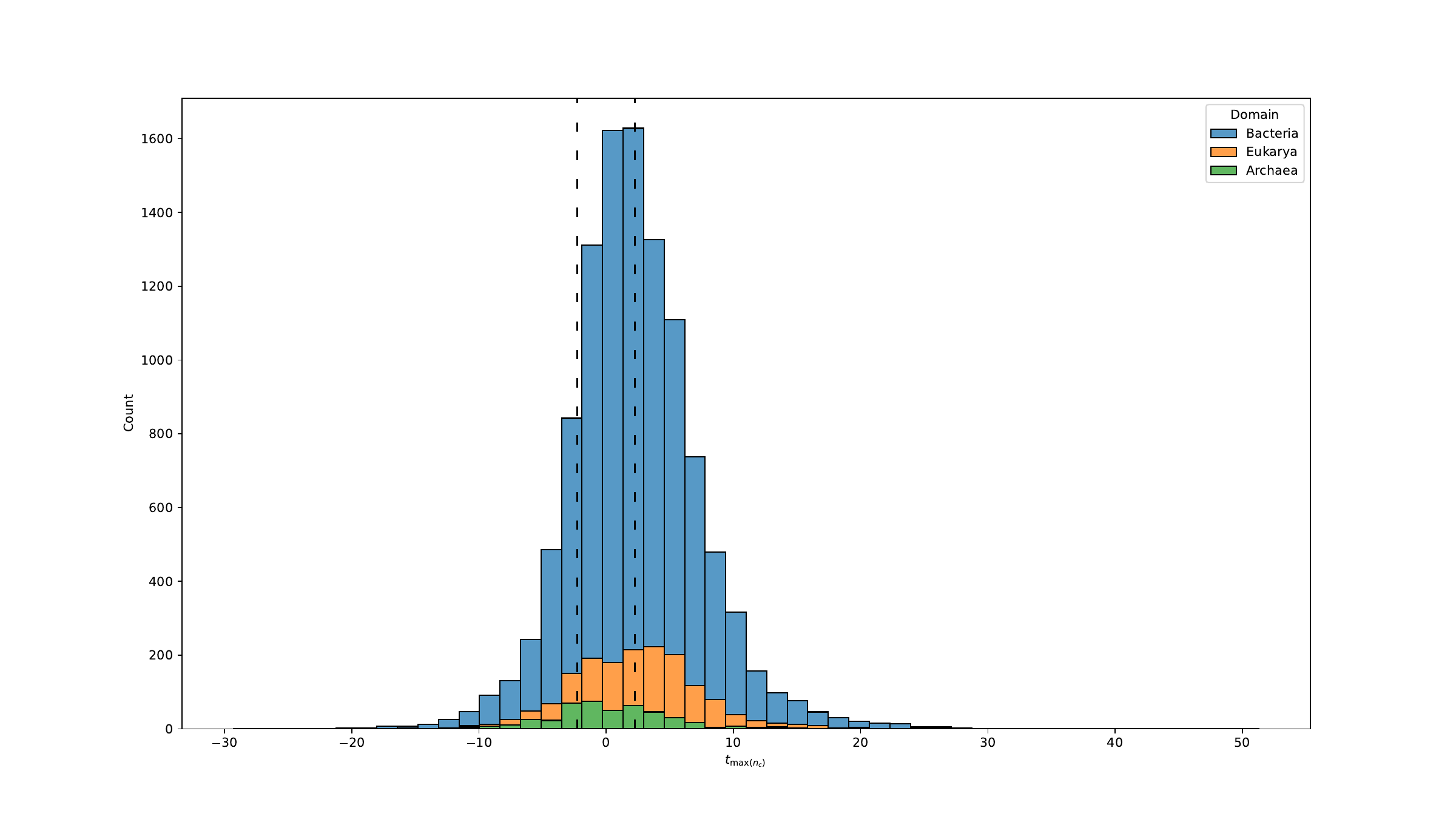}}
     \subfloat[Fraction of $p<0.05$ per domain.]{\includegraphics[width=0.41\textwidth]{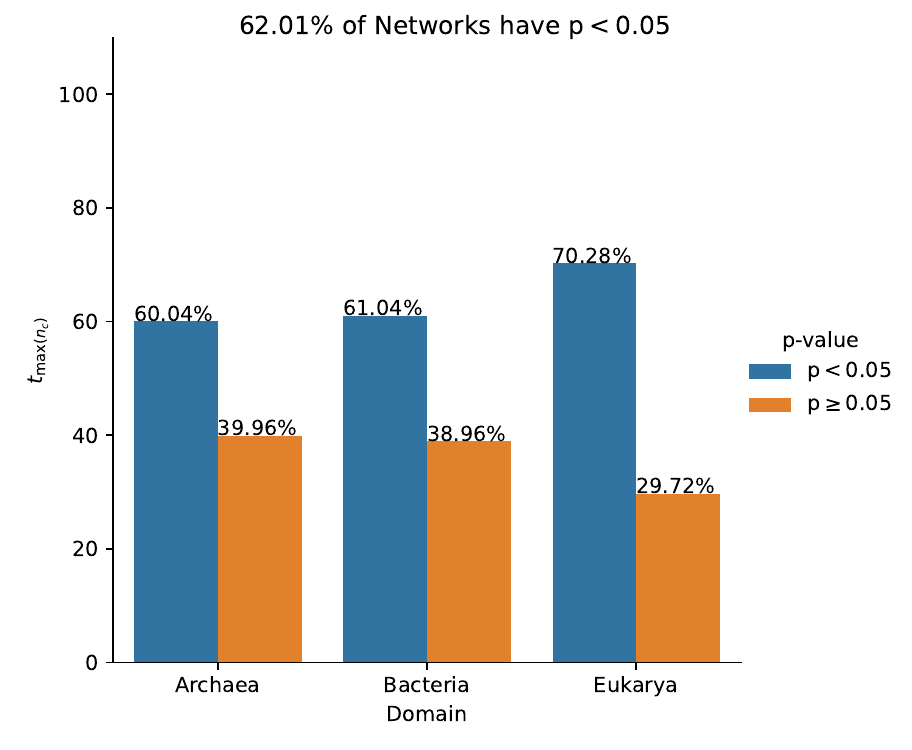}}    } 
\caption{Histograms displaying $t_p$ for the community structure parameters of the metabolic networks. As before, we choose to stack the distributions for better visibility. On the right side, we show the percentage of networks with p$<0.05$ per domain.}
    \label{fig:t_percent_surp_2}
\end{figure}

\begin{figure}[htbp]

\makebox[\linewidth][c]{
    \centering
     
\subfloat[Pielou's Index ($t_{PI}$).]{\includegraphics[width=0.59\textwidth]{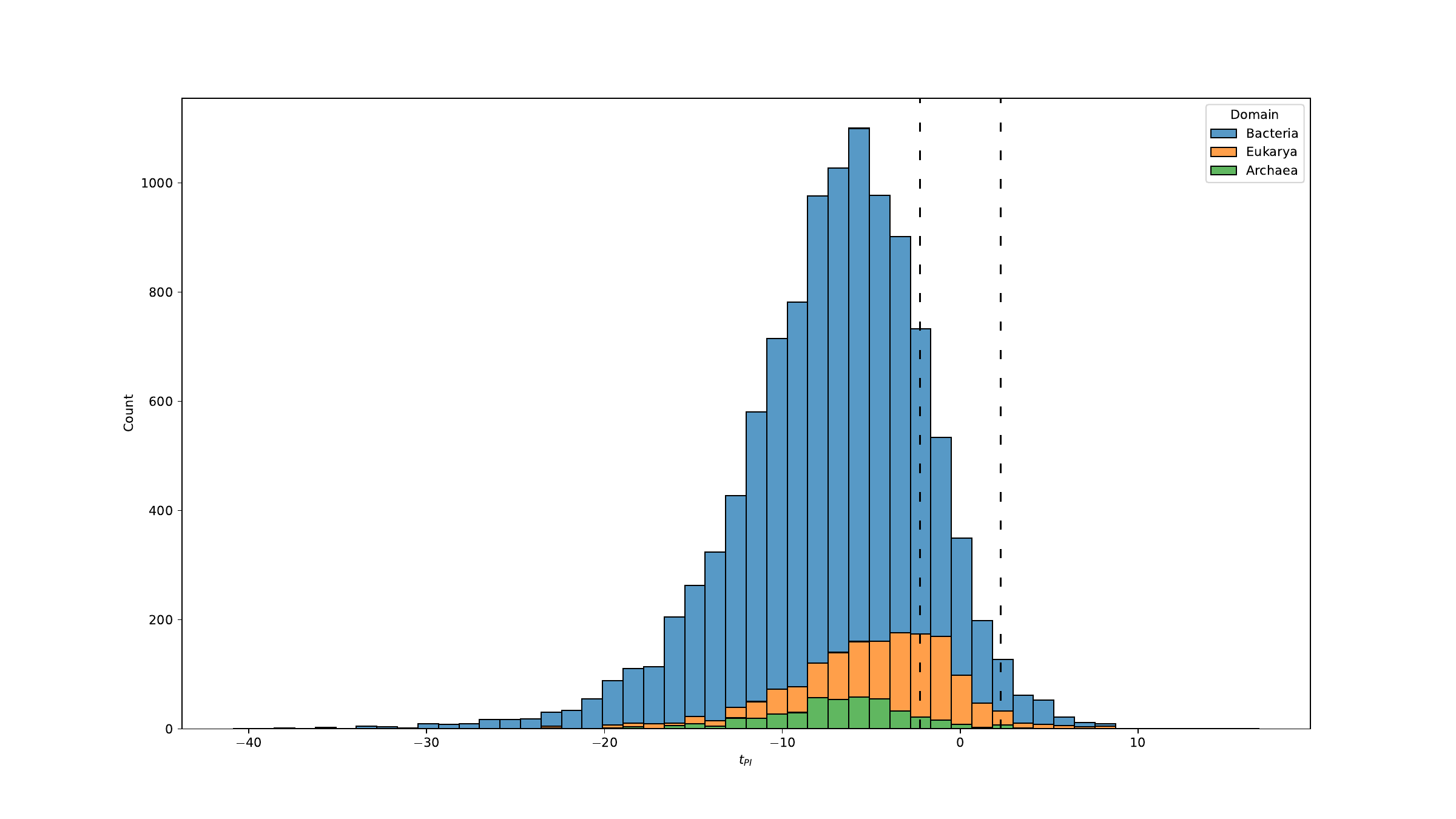}}
     \subfloat[Fraction p$<0.05$ per domain.]{\includegraphics[width=0.41\textwidth]{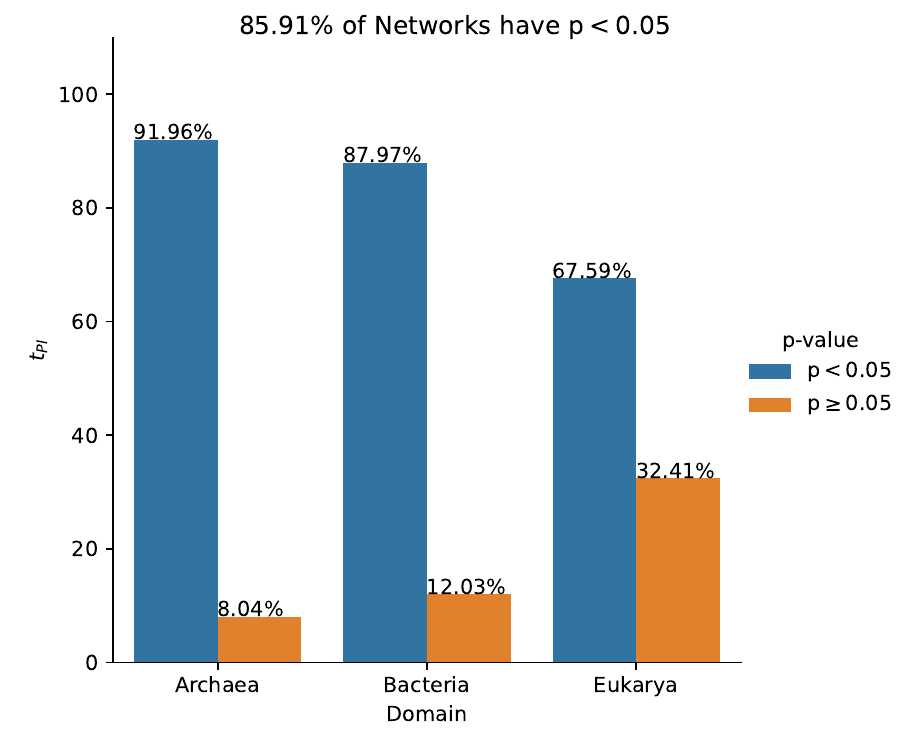}}    } 
 
\caption{Histograms displaying $t_{PI}$ for Pielou's index of the metabolic networks. Once again, we choose to stack the distributions for better visibility. On the right side, we show the percentage of networks with p$<0.05$ per domain.}
    \label{fig:t_percent_surp_3}
\end{figure}

Figures \ref{fig:t_percent_surp_2} and \ref{fig:t_percent_surp_3} show the rest of the statistics related to the community structure of the networks. {We note that the difference between real networks and randomized ones is most significant in terms of the Pielou's index, with $85.91\%$ of them exhibiting statistically significant deviations from their randomized counterparts. Among networks with a significant deviation for the Pielou's index, $t_{PI}$, there is a clear tendency towards more even community distributions: \(83.79\%\) of them have \(t_{PI} \leq 0\), indicating a higher than expected Pielou's index and, consequently, more evenness in the community distribution. This trend is less pronounced among eukaryotes, where $32.04\%$ of the networks do not show a difference that is statistically significant and \(64.06\%\) of the networks have \(t_{PI} \leq 0\).}

{In the other end of the spectrum, we see that the number of communities is higher for $65.87\%$ of metabolic networks and that the average community size is smaller for $65.99\%$ of organisms. As the number of nodes is constant, those two parameters must be correlated. If the average community size is smaller, then the number of communities must be higher. Indeed, $99.56\%$ of organisms that have a larger number of communities  also have a smaller average community size. Only $0.44\%$ of these organisms show a result that is statistically significant for one of the characteristics, but not for the other.} 

{In relation to the maximum community size, approximately half of the organisms ($46.63\%$) show a smaller maximum community size than expected. However, this pattern is not as widespread, as $37.99\%$ of organisms show no statistically significant deviation from their randomized counterparts. At the kingdom level, a distinct behavior emerges within the Animalia kingdom, where 
$79.47\%$ of organisms display a smaller maximum community size, while only 
$4.53\%$ show a larger one and 
$16.00\%$ present no significant difference. This pattern among animals may reflect a greater degree of metabolic specialization in more complex multicellular organisms.}

Another characteristic we examined was the presence of what we call 'bridges'. Due to the degeneracy of {the} Surprise {parameter, one observes} that different partitions may result in the same maximum Surprise value. In other words, certain nodes or groups of nodes can be assigned to different communities and still yield the same Surprise. We label these nodes as 'bridges' because they facilitate connections between different communities and can belong to either of them. Currently, we are not concerned with the biological significance of such bridges. Instead, our focus is on comparing their presence in real metabolic networks with their presence in their randomized versions. Figures \ref{fig:t_percent_bridge}-a and -b illustrate the distribution of $t_{N_{B}}$ associated with the number of bridges ($N_{B}$) identified in the metabolic networks. The occurrence of bridges is less common in the vast majority of real networks, with 95.42\% of them possessing fewer bridges than their randomized counterparts. In real-world networks, it is more probable that each metabolite has an unambiguous community classification, indicating a higher degree of specialization.

\begin{figure}[htbp]

\makebox[\linewidth][c]{
    \centering
     
\subfloat[Number of Bridges ($t_{N_B}$).]{\includegraphics[width=0.59\textwidth]{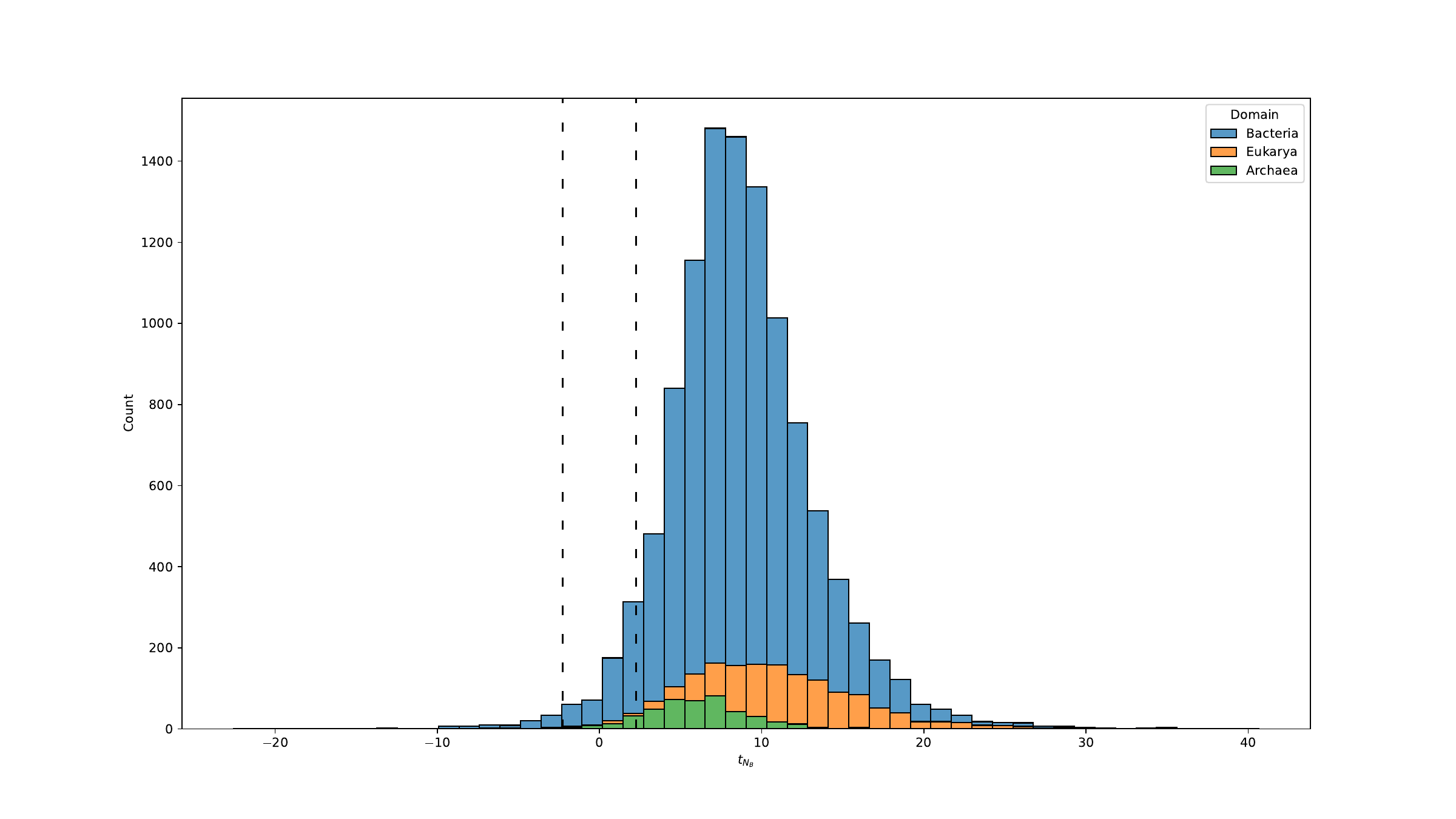}}
     \subfloat[Fraction p$<0.05$ per domain.]{\includegraphics[width=0.41\textwidth]{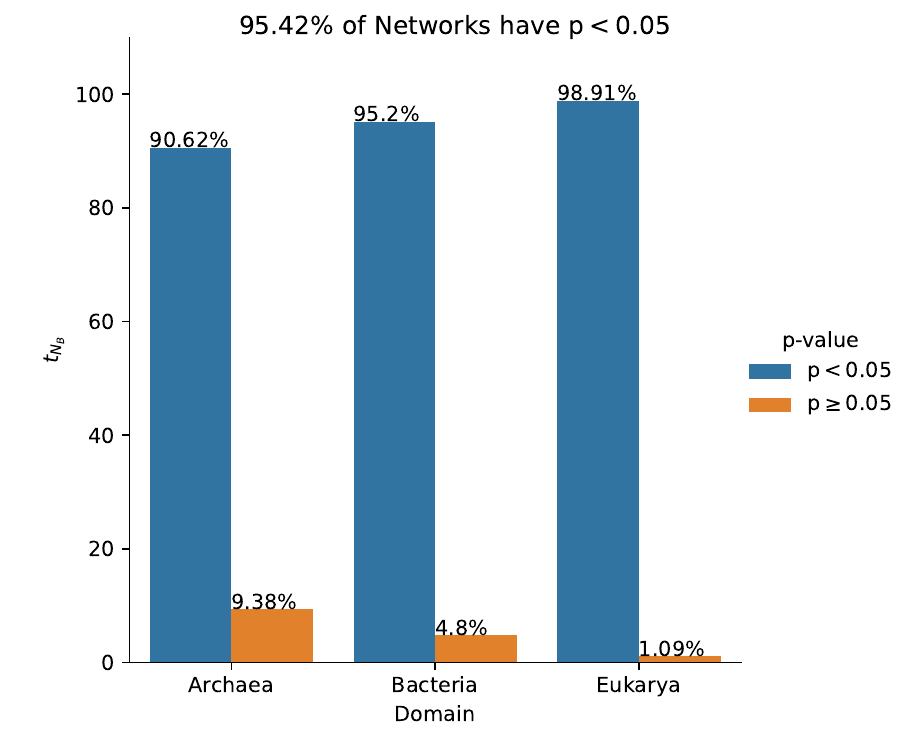}}    } 
 
\caption{Histograms displaying $t_{N_B}$ for the number of bridges in metabolic networks. Once again, we choose to stack the distributions for better visibility. In the right side we show the percentage of networks with p$<0.05$ per domain.}
    \label{fig:t_percent_bridge}
\end{figure}

\section{Conclusion}\label{sec:conc}

We introduced a new method for constructing metabolic networks that more accurately reflects the transformation of each molecule in a biochemical reaction. This approach not only improves the representation of metabolic processes but also effectively addresses the challenge of properly managing currency metabolites within the network. 

Using this novel approach we created and analyzed the metabolic networks of 10912 different organisms spanning the three taxonomic domains: Eukarya, Bacteria and Archaea. Our analysis revealed that all metabolic networks are disassortative. Moreover, metabolic networks exhibit a higher average local clustering coefficient compared to their global clustering coefficient. This lack of correlation between local and global clustering is a common feature observed in real-world networks across various systems and aligns with findings from previous studies\cite{gamermann2, newman2003structure, estrada}. 

{Early studies on metabolic network topology suggested that these systems exhibit scale-free properties \cite{barabasi2004network}. However, more recent statistical analyses have questioned the prevalence of true power-law behavior in metabolic networks\cite{gamermann2, broido2019scale}. In this context, our findings support the latter perspective. This is particularly evident in the scarcity of high-degree nodes or hubs, which is consistent across all three taxonomic domains.} {This feature is likely due to our more rigorous handling of currency metabolites. These ubiquitous molecules, which typically form large hubs in substrate networks, tend to artificially inflate node degree and alter the network topology \cite{takemoto2014metabolic,ravasz2002hierarchical,ma2003reconstruction}. By restricting such metabolites to biologically meaningful connections, we were able to mitigate these distortions, leading to a more accurate reflection of the networks true structure.}

{A direct consequence of this refinement is an increase in both the average distance and assortativity of the networks. This observation is consistent with previous studies showing that the inclusion of currency metabolites can distort estimates of the distance between nodes in the network \cite{ma2003reconstruction}.} {Further insight can be gained by comparing our results with those from our earlier work based on traditional substrate networks \cite{gamermann2}. While the present networks exhibit a 55.3\% increase in the number of nodes, they simultaneously show a
16\% decrease in the number of links. This discrepancy suggests that a substantial
portion of the previously identified connections may lack biochemical relevance. In line with this, the average degree also shows a substantial decrease of 54\% (from 5.69 to 3.09). Taken together, these results imply that metabolic networks are less interconnected than previously thought, which could have significant implications for our understanding of metabolic processes.}

Beyond topological properties, we also investigated the community structure of these metabolic networks. Our analysis shows that the networks are composed of numerous small communities, with an average size of 
$\bar{n}_c = 3.64 \pm 0.20$
nodes and an average community count of 
$N_c = 381.15 \pm 118.16.$
Moreover, the communities are relatively uniform in size, as reflected by the average Pielou's index, which is close to one 
$PI = 0.961 \pm 0.004.$.

Next, we compared the real metabolic networks with randomized versions that maintained the same degree distribution. This comparison aimed to determine whether the observed network properties and community structures were solely a consequence of the degree distribution or if they resulted from different evolutionary pressures. We observe that the community structure of real metabolic networks significantly deviates from their randomized versions. Employing a student t-test with a confidence level of 95\% (p-value $\leq 0.05$), we discovered that 99.98\% of real metabolic networks exhibit a higher Surprise value than their randomized counterparts. This suggests that real-world networks possess a more pronounced community structure, as the Surprise function quantifies the level of surprise (or unlikeliness) in obtaining, at random, a partition with a similar enrichment of intra-community links. Further support for this conclusion comes from the observation that real metabolic networks exhibit a higher local clustering coefficient (99.38\% of networks) and a greater fraction of intra-community links (98.78\% of networks) compared to their randomized versions.

The most significant deviations between the topological properties of real metabolic networks and their randomized counterparts occur in the local and global clustering coefficients. For over 99\% of the networks examined, these differences are statistically significant, indicating that real metabolic networks exhibit higher clustering coefficients than their randomized versions. {In addition to this, $96.67\%$ of organisms exhibit a lower-than-expected average distance compared to their randomized counterparts, even though this behavior is less pronounced in members of {the} kingdom Animalia.} 

In summary, our study of metabolic networks points to two main conclusions. First, it reveals that the prevalent method in the literature of linking every product to each substrate in chemical reactions often leads to metabolic networks in which many connections are biologically irrelevant. This approach can result in networks that do not accurately reflect the true interactions and pathways within biological systems. Our method is an improvement in this sense as every link has a biochemical significance in our networks.

{Second, we demonstrate that the community structure of metabolic networks arises not solely from their degree distribution, but rather as a consequence of a multitude of factors, including specialization, regulatory mechanisms, and the evolutionary history of the networks}. This emphasizes the need for evolutionary models to incorporate additional mechanisms beyond the replication of the degree distribution, which is usually the goal in different models.

\appendix

\section{Algorithm Explanation and examples} \label{appendixA}

In this appendix we explain in more detail the algorithm used to systematically obtain the relevant network links for each graph representing an organism's metabolic network from the chemical reactions identified as possibly taking place in its cells. For that we will follow examples of reactions where the algorithm is applied step by step.

We want to create a network that connects species by identifying how different molecules combine to form more complex ones or how they breakdown into smaller pieces (what is being transformed into what?). The relevant data used can be found in the KEGG database which has two kinds of molecules that appear in the chemical reactions: compounds, described by identifiers like \texttt{CXXXXX} and glycans, described by identifiers of the kind \texttt{GXXXXX}, where the \texttt{X}'s are integer digits. Most molecules are long chains of carbon, or multiple connected chains and/or rings, but we are going to focus on the chemical composition of the molecules and not their shapes. In fact, some reactions are just rearrangements in the shape of a molecule. Most works in the field start by connecting all substrates appearing in a reaction (left-hand side) to all products (right-hand side). Now we are going to connect one side of the equation to the other reconstructing each substrate by merging the products together. To do that, we follow the steps:

\bi
\item Define the left side of the equation as the side with the least number of different molecules (our focus is undirected networks, so the choice of product/substrate is arbitrary).
\item {Get the chemical formula for each metabolite in the equation, when possible.}
\item Compare each metabolite formula with each possible combination of the formulas for the metabolites in the other side by creating a score that penalizes differences (different amounts of atoms in the formulas being compared).
\item {Connect the metabolite on the left to all metabolites on the right that better reconstruct it (smallest difference) and eliminate them from the equation.}
\item {Repeat the process until the left side of the equation is empty.}
\item {For compounds where it is not possible to reconstruct their chemical formulas, compare their names.}
\item {If the left side could not be defined as the same number of compounds is the same on both sides, the process is done twice with each side as the starting one and the best combinations are chosen.}
\ei

Let's focus on examples. In Scheme \ref{sc1} we show the reaction where an oxygen molecule combines with two hydrogen molecules to form two water molecules. The scheme shows the reaction, with the chemical compositions followed then by the table it uses to evaluate how the left hand side of the equation (defined by the algorithm as the side with the water molecules, since it had the least number of different molecules) is better reconstructed (compared) with every possible combination of the molecules in the right hand side.

\begin{schemea}[h]
\begin{framed}
\bc \underline{Water Formation}: $\underbrace{\textrm{Oxygen}}_{O_{2}} + 2 \underbrace{\textrm{Hydrogen}}_{H_{2}} \leftrightarrow 2 \underbrace{\textrm{H2O}}_{H_{2}O_{}}$ \ec
\bc
\scalebox{1.0}{\bt{|c|ccccc|}
\hline
 & $O_{2}$ & $H_{2}$ & 2 $H_{2}$ & $O_{2}$ + $H_{2}$ & $O_{2}$ + 2 $H_{2}$  \\
\hline$H_{2}O_{}$ & 0.37500 & 0.71429 & 0.75000 & 0.29412 & 0.33333 \\
2 $H_{2}O_{}$ & 0.09091 & 0.84615 & 0.71429 & 0.04348 & 0.00000 \\
\hline
\et}
\ec
\bc
\scalebox{1.0}{\begin{tikzpicture}
\node[draw, red] (S1) at (0, 0) {  $2 \underbrace{\textrm{H2O}}_{H_{2}O_{}}$ };
\node (A) [right=0.1 of S1] {$\Leftarrow$};
\node[draw, red] (P1) [right=0.1 of A] {  $\underbrace{\textrm{Oxygen}}_{O_{2}}$ };
\node (plus2) [right=.1 of P1] {+};
\node[draw, red] (P2) [right=.1 of plus2] {  $2 \underbrace{\textrm{Hydrogen}}_{H_{2}}$ };
\node[draw, red] (V1) [below=1 of S1] {0.0};
\draw[red] (S1) -- (V1);
\draw[red] (P1) -- ($ (P1) - (0, 1) $);
\draw[red] ($ (P1) - (0, 1) $) -- (V1);
\draw[red] (P2) -- ($ (P2) - (0, 1) $);
\draw[red] ($ (P2) - (0, 1) $) -- (V1);
\end{tikzpicture}}

This reaction is balanced! (zero points)
\ec
\end{framed}
\caption{Scheme for water formation reaction. First the reaction is presented, below the table with the comparisons and finally the connections shown in the reaction.}\label{sc1}
\end{schemea}

Two water molecules were perfectly reconstructed by one oxygen plus two hydrogen molecules. Note {that} in the original reaction, water was on the right hand side, but the algorithm reverses the reaction, though the arrow in the picture with the links indicates the original direction the reaction was fed to the algorithm. The final zero linking all compounds is the difference between the reconstructed molecule (two waters) and the molecules it was reconstructed with (zero means no difference). The numbers in the table are the differences (the score) between either one or two water molecules (left hand side of the equation) with all possible combinations of the molecules on the right hand side. This score is evaluated as follows: except for carbon, hydrogen and electrons, each chemical element counts one point, while carbon counts 2 and hydrogen and electrons count $\frac{1}{10}$. These values were chosen given the problem at hand: reconstruction of mainly organic molecules whose main element is carbon and where hydrogen or electrons are of small consequence and can be easily gained or lost. In comparing two groups of elements (a molecule in the left hand side with a combination of molecules in the right) means summing up all differences and dividing by the total sum of each group. For example, the total sum for a water molecule ($H_2 O$) is 1.2, and the total sum for a oxygen molecule plus a hydrogen molecule ($O_2 + H_2$) is 2.2. The differences between these two groups is 1 (a single oxygen atom) such that the final score for this comparison, shown in the table, is $\frac{1}{1.2+2.2}=0.29412$, as can be seen in the first line, forth row of numbers in the table.

Given the table, the algorithm searches the smallest number and links the group in the line with the corresponding group in the row. In this case, would create two links: one between water and oxygen and one between water and hydrogen. In a more complex reaction (as will be shown next), the identified combination is removed from the equation and a new table is created with the still remaining molecules in the reaction. In this present case though, two water molecules are reconstructed from a single oxygen molecule combined with two hydrogen ones and nothing is left (the smallest number in the first table is a zero, already).

Now a side note: the number of columns in the table is given by the number of molecules in the right hand side. Each molecule in the right hand side can or cannot be present in a possible reconstruction. So, considering $c_i$ for $i$=1, 2, ..., $N$, the stoichiometry coefficient of each of the $N$ different molecules in the right hand side, one will have $\left(\displaystyle\prod_{i=1}^{N}(c_i+1)\right)-1$ possible combinations where each of the molecules in the right hand side can contribute with either 0, 1, ..., $c_i$ molecules in the combination (the $c_i+1$ factor) and the minus one is removing the combination where all compounds in the right hand side contribute with 0 molecules for the combination. In this first example the right hand side had one oxygen ($c_1=1$) and two hydrogen ($c_2=2$), so $2\times 3-1=5$ possible groupings appear in the table.  

Now let's see the Scheme \ref{sc2} where a fictitious reaction reconstructs water with the help of NAD. This reaction does not exist, but illustrates a case when after a first reconstruction a second one happens.

\begin{schemea}[h]
\begin{framed}
\bc \underline{Fictious Water Formation}: $2 \underbrace{\textrm{H2O}}_{H_{2}O_{}} + 2 \underbrace{\textrm{NADH}}_{C_{21}H_{29}N_{7}O_{14}P_{2}} \leftrightarrow \underbrace{\textrm{Oxygen}}_{O_{2}} + 3 \underbrace{\textrm{Hydrogen}}_{H_{2}} + 2 \underbrace{\textrm{NAD+}}_{C_{21}H_{28}N_{7}O_{14}P_{2}}$ \ec
\bc
\scalebox{0.15}{\bt{|c|ccccccccccccccccccccccc|}
\hline
 & $O_{2}$ & $H_{2}$ & 2 $H_{2}$ & 3 $H_{2}$ & $O_{2}$ + $H_{2}$ & $O_{2}$ + 2 $H_{2}$ & $O_{2}$ + 3 $H_{2}$ & $C_{21}H_{28}N_{7}O_{14}P_{2}$ & 2 $C_{21}H_{28}N_{7}O_{14}P_{2}$ & $O_{2}$ + $C_{21}H_{28}N_{7}O_{14}P_{2}$ & $O_{2}$ + 2 $C_{21}H_{28}N_{7}O_{14}P_{2}$ & $H_{2}$ + $C_{21}H_{28}N_{7}O_{14}P_{2}$ & 2 $H_{2}$ + $C_{21}H_{28}N_{7}O_{14}P_{2}$ & 3 $H_{2}$ + $C_{21}H_{28}N_{7}O_{14}P_{2}$ & $H_{2}$ + 2 $C_{21}H_{28}N_{7}O_{14}P_{2}$ & 2 $H_{2}$ + 2 $C_{21}H_{28}N_{7}O_{14}P_{2}$ & 3 $H_{2}$ + 2 $C_{21}H_{28}N_{7}O_{14}P_{2}$ & $O_{2}$ + $H_{2}$ + $C_{21}H_{28}N_{7}O_{14}P_{2}$ & $O_{2}$ + 2 $H_{2}$ + $C_{21}H_{28}N_{7}O_{14}P_{2}$ & $O_{2}$ + 3 $H_{2}$ + $C_{21}H_{28}N_{7}O_{14}P_{2}$ & $O_{2}$ + $H_{2}$ + 2 $C_{21}H_{28}N_{7}O_{14}P_{2}$ & $O_{2}$ + 2 $H_{2}$ + 2 $C_{21}H_{28}N_{7}O_{14}P_{2}$ & $O_{2}$ + 3 $H_{2}$ + 2 $C_{21}H_{28}N_{7}O_{14}P_{2}$  \\
\hline$C_{21}H_{29}N_{7}O_{14}P_{2}$ & 0.94278 & 0.99413 & 0.98829 & 0.98248 & 0.93723 & 0.93172 & 0.92624 & 0.00074 & 0.33268 & 0.01525 & 0.33917 & 0.00074 & 0.00220 & 0.00367 & 0.33333 & 0.33399 & 0.33464 & 0.01523 & 0.01665 & 0.01808 & 0.33982 & 0.34046 & 0.34110 \\
2 $C_{21}H_{29}N_{7}O_{14}P_{2}$ & 0.97097 & 0.99706 & 0.99413 & 0.99120 & 0.96812 & 0.96527 & 0.96243 & 0.33399 & 0.00074 & 0.32101 & 0.00805 & 0.33268 & 0.33137 & 0.33007 & 0.00000 & 0.00074 & 0.00147 & 0.31973 & 0.31845 & 0.31717 & 0.00731 & 0.00804 & 0.00876 \\
$H_{2}O_{}$ & 0.37500 & 0.71429 & 0.75000 & 0.77778 & 0.29412 & 0.33333 & 0.36842 & 0.96522 & 0.98246 & 0.96620 & 0.98271 & 0.96532 & 0.96542 & 0.96552 & 0.98248 & 0.98251 & 0.98253 & 0.96629 & 0.96639 & 0.96648 & 0.98273 & 0.98276 & 0.98278 \\
2 $H_{2}O_{}$ & 0.09091 & 0.84615 & 0.71429 & 0.73333 & 0.04348 & 0.00000 & 0.04000 & 0.93162 & 0.96522 & 0.93352 & 0.96571 & 0.93182 & 0.93201 & 0.93220 & 0.96527 & 0.96532 & 0.96537 & 0.93370 & 0.93388 & 0.93407 & 0.96576 & 0.96581 & 0.96586 \\
\hline
\et}
\ec
\bc
\scalebox{1.0}{\bt{|c||ccccc|}
\hline & $O_{2}$ & $H_{2}$ & 2 $H_{2}$ & $O_{2}$ + $H_{2}$ & $O_{2}$ + 2 $H_{2}$  \\
\hline$H_{2}O_{}$ & 0.37500 & 0.71429 & 0.75000 & 0.29412 & 0.33333 \\
2 $H_{2}O_{}$ & 0.09091 & 0.84615 & 0.71429 & 0.04348 & 0.00000 \\
\hline
\et}
\ec
\bc
\scalebox{0.9}{\begin{tikzpicture}
\node[draw, red] (S1) at (0, 0) {  $2 \underbrace{\textrm{H2O}}_{H_{2}O_{}}$ };
\node (plus1) [right=.1 of S1] {+};
\node[draw, blue] (S2) [right=.1 of plus1] {  $2 \underbrace{\textrm{NADH}}_{C_{21}H_{29}N_{7}O_{14}P_{2}}$ };
\node (A) [right=0.1 of S2] {$\Rightarrow$};
\node[draw, red] (P1) [right=0.1 of A] {  $\underbrace{\textrm{Oxygen}}_{O_{2}}$ };
\node (plus3) [right=.1 of P1] {+};
\node[draw, black] (P2) [right=.1 of plus3] {  $3 \underbrace{\textrm{Hydrogen}}_{H_{2}}$ };
\node (plus4) [right=.1 of P2] {+};
\node[draw, blue] (P3) [right=.1 of plus4] {  $2 \underbrace{\textrm{NAD+}}_{C_{21}H_{28}N_{7}O_{14}P_{2}}$ };
\node[draw, blue] (V2) [below=1 of S2] {0.0};
\draw[blue] (S2) -- (V2);
\draw[black] (P2) -- ($ (P2) - (0, 1) $);
\draw[black] ($ (P2) - (0, 1) $) -- (V2);
\draw[blue] (P3) -- ($ (P3) - (0, 1) $);
\draw[blue] ($ (P3) - (0, 1) $) -- (V2);
\node[draw, red] (V1) [below=2 of S1] {0.0};
\draw[red] (S1) -- (V1);
\draw[red] (P1) -- ($ (P1) - (0, 2) $);
\draw[red] ($ (P1) - (0, 2) $) -- (V1);
\draw[black] (P2) -- ($ (P2) - (0, 2) $);
\draw[black] ($ (P2) - (0, 2) $) -- (V1);
\end{tikzpicture}}

This reaction is balanced! (zero points)
\ec
\end{framed}
\caption{Fictitious water formation involving NAD.}\label{sc2}
\end{schemea}

Now the first table is barely readable since it has many combinations. The right hand side has now $c_i$'s equal to 1, 2 and 3, so we have a total of $2\times3\times4-1=23$ columns in this table (please make a zoom in the \texttt{PDF} to look the table and find the zero, it happens for the column comparing two NADH molecules with the two NAD+ plus one hydrogen). There is {a} second zero, for the two water molecules combining with the oxygen plus two hydrogen. When two equal minima occur (two zeros in this case), it prioritizes reconstructing the bigger molecule first. So it removes the two molecules of NADH, the two molecules of NAD+ and one molecule of hydrogen and proceeds to execute the algorithm with the remaining molecules, which are now the same ones as in the first example, therefore the second table is equal to the one in the Scheme \ref{sc1}. The final result for this example is water connected only with oxygen and hydrogen, while NADH connects with hydrogen and NAD+.

Again we see that all substrates were completely reconstructed by joining the products. While other reconstructed molecules have a color identifying their relations, the hydrogen appears in a black square because it cannot be unequivocally associated with a single molecule in the left hand side: One hydrogen molecule is broken to reconstruct the 2 NADH molecules from the two NAD+ and the other two hydrogen molecules are combined with the oxygen to build two water molecules.

Note that from this algorithm one directly has a way to check if a reaction is balanced. Indeed the schemes mention it and point out that the final comparisons result in zero points. If one makes the comparison of the grouping of all molecules in one side of the equation with the molecules in the other side, one should get a difference of zero, if the reaction is balanced (mass conservation). {We} have a total of 12199 reactions downloaded from the KEGG database, from these, 9509 (77.949\%) have zero points when the comparison is made. Also there are 192 reactions where the same compound is found in both sides of the equation. From these 99 are not balanced, meaning that in 93 of these the same compound issue can be dealt with by setting some details. For example reaction R02185:

\bc$\underbrace{\textrm{(Polyphosphate)$_{n+1}$}}_{H_{18}O_{49}P_{16}} + \underbrace{\textrm{D-Glucose}}_{C_{6}H_{12}O_{6}} \leftrightarrow \underbrace{\textrm{(Polyphosphate)$_{n}$}}_{H_{17}O_{46}P_{15}} + \underbrace{\textrm{D-Glucose 6-phosphate}}_{C_{6}H_{13}O_{9}P_{}}$\ec

The polyphosphate (C00404) appears on both sides. Its formula in KEGG is $H_4P_2O_7(HPO_3)_n$ , so we can see that the ``poly'' is the $n$ (how many phosphates) are in the big molecule. So we can adjust the mass conservation by setting the $n$ in the left hand side to $n+1$. The algorithm, if not forced otherwise, will always substitute an unknown number like $n$ by 13. But now we can see the result of running the algorithm for this reaction in Scheme \ref{sc3}. In this case it will deal differently with the $n$. When an undermined number like $n$ is found, the algorithm by default substitutes it by 13 (user's choice).

\begin{schemea}[h]
\begin{framed}
\bc \underline{R02185}: $\underbrace{\textrm{(Polyphosphate)$_{n+1}$}}_{H_{18}O_{49}P_{16}} + \underbrace{\textrm{D-Glucose}}_{C_{6}H_{12}O_{6}} \leftrightarrow \underbrace{\textrm{(Polyphosphate)$_{n}$}}_{H_{17}O_{46}P_{15}} + \underbrace{\textrm{D-Glucose 6-phosphate}}_{C_{6}H_{13}O_{9}P_{}}$ \ec
\bc
\scalebox{1.0}{\bt{|c|c|}
\hline
 & $C_{6}H_{13}O_{9}P_{}$  \\
\hline$C_{6}H_{12}O_{6}$ & 0.09647 \\
$H_{}O_{3}P_{}$ & 0.70073 \\
\hline
\et}
\ec
\bc
\scalebox{1.0}{\bt{|c|||}
\hline \\
\hline\hline
\et}
\ec
\bc
\scalebox{1.0}{\bt{|c||ccc|}
\hline & $H_{}O_{3}P_{}$ & $C_{6}H_{12}O_{6}$ & $H_{}O_{3}P_{}$ + $C_{6}H_{12}O_{6}$  \\
\hline$C_{6}H_{13}O_{9}P_{}$ & 0.70073 & 0.09647 & 0.00000 \\
\hline
\et}
\ec
\bc
\scalebox{0.8}{\begin{tikzpicture}
\node[draw, red] (S1) at (0, 0) {  $\underbrace{\textrm{(Polyphosphate)$_{n+1}$}}_{H_{18}O_{49}P_{16}}$ };
\node (plus1) [right=.1 of S1] {+};
\node[draw, blue] (S2) [right=.1 of plus1] {  $\underbrace{\textrm{D-Glucose}}_{C_{6}H_{12}O_{6}}$ };
\node (A) [right=0.1 of S2] {$\Rightarrow$};
\node[draw, black] (P1) [right=0.1 of A] {  $\underbrace{\textrm{(Polyphosphate)$_{n}$}}_{H_{17}O_{46}P_{15}}$ };
\node (plus3) [right=.1 of P1] {+};
\node[draw, black] (P2) [right=.1 of plus3] {  $\underbrace{\textrm{D-Glucose 6-phosphate}}_{C_{6}H_{13}O_{9}P_{}}$ };
\node[draw, blue] (V2) [below=1 of S2] {0.09647058823529411};
\draw[blue] (S2) -- (V2);
\draw[black] (P2) -- ($ (P2) - (0, 1) $);
\draw[black] ($ (P2) - (0, 1) $) -- (V2);
\node[draw, red] (V1) [below=2 of S1] {0.0};
\draw[red] (S1) -- (V1);
\draw[black] (P2) -- ($ (P2) - (0, 2) $);
\draw[black] ($ (P2) - (0, 2) $) -- (V1);
\end{tikzpicture}}

This reaction is balanced! (zero points)
\ec
\end{framed}
\caption{Reaction with the same metabolite in both sides.}\label{sc3}
\end{schemea}

Now, when printing the reaction, it uses 13 in place of $n$, but in the tables, instead of the formulas we see in the equation for the polyphosphate, we encounter $HO_3P$ in the left hand side and the other polyphosphate has been set to $H_0O_0P_0$ (it is nowhere to be seen). What happens here is that when it identifies the same compound in both sides, it removes what is equal to both until it is nonzero only in one side. Another thing to note here about the algorithm is how it works when it cannot differentiate between the left- and right-hand sides (both sides in this case have two different compounds) because the number of different molecules is the same in both sides. In this case, it runs the algorithm twice, each time considering a different side as the starting (left) one. That's why we see two tables, in the first D-Glucose and polyphosphate were considered left hand-side and in the second table D-Glucose 6-phosphate and polyphosphate (actually nothing now) were. From the two tables it chooses the best reconstruction (the one resulting in a final zero).

In Scheme \ref{sc5} we show reaction R05740, which has a compound that cannot be broken down into chemical elements, the Ferrocytochrome. This protein complex participates in the reaction giving and receiving either electrons or protons. Now the first table has 35 columns where Crepenyate and Linoleate are connected and the remaining molecules in the right hand side make 17 different groupings to be compared with each of the four possibilities for the left hand side. Then the Ferrocytochromes are connected and only the water molecules remain in the left hand side to be reconstructed. In this case, what was compared in the last step was the name of the compounds (Ferrocytochrome or Ferricytochrome) between them but, when comparing this with a chemical element, since they are obviously not the same, a penalty (10 points) is increased in the difference. Note that this reaction appears as unbalanced (it is not among the 77.95\% of the reactions mentioned above), since it has a final punctuation of 0.000612. This final discrepancy is the result of comparing the names between the two cytochromes, which are actually the same protein complex, but have received and donated electrons for the molecules transformed in the reaction.

\begin{schemea}[h]
\begin{framed}
\bc \underline{R05740}: $\underbrace{\textrm{Linoleate}}_{C_{18}H_{32}O_{2}} + 2 \underbrace{\textrm{Ferrocytochrome b5}}_{Ferrocytochrome b5_{}} + \underbrace{\textrm{Oxygen}}_{O_{2}} + 2 \underbrace{\textrm{H+}}_{H_{}} \leftrightarrow \underbrace{\textrm{Crepenynate}}_{C_{18}H_{30}O_{2}} + 2 \underbrace{\textrm{Ferricytochrome b5}}_{Ferricytochrome b5_{}} + 2 \underbrace{\textrm{H2O}}_{H_{2}O_{}}$ \ec
\bc
\scalebox{0.08}{\bt{|c|ccccccccccccccccccccccccccccccccccc|}
\hline
 & $C_{18}H_{32}O_{2}$ & $Ferrocytochrome b5_{}$ & 2 $Ferrocytochrome b5_{}$ & $C_{18}H_{32}O_{2}$ + $Ferrocytochrome b5_{}$ & $C_{18}H_{32}O_{2}$ + 2 $Ferrocytochrome b5_{}$ & $O_{2}$ & $C_{18}H_{32}O_{2}$ + $O_{2}$ & $Ferrocytochrome b5_{}$ + $O_{2}$ & 2 $Ferrocytochrome b5_{}$ + $O_{2}$ & $C_{18}H_{32}O_{2}$ + $Ferrocytochrome b5_{}$ + $O_{2}$ & $C_{18}H_{32}O_{2}$ + 2 $Ferrocytochrome b5_{}$ + $O_{2}$ & $H_{}$ & 2 $H_{}$ & $C_{18}H_{32}O_{2}$ + $H_{}$ & $C_{18}H_{32}O_{2}$ + 2 $H_{}$ & $Ferrocytochrome b5_{}$ + $H_{}$ & 2 $Ferrocytochrome b5_{}$ + $H_{}$ & $Ferrocytochrome b5_{}$ + 2 $H_{}$ & 2 $Ferrocytochrome b5_{}$ + 2 $H_{}$ & $C_{18}H_{32}O_{2}$ + $Ferrocytochrome b5_{}$ + $H_{}$ & $C_{18}H_{32}O_{2}$ + 2 $Ferrocytochrome b5_{}$ + $H_{}$ & $C_{18}H_{32}O_{2}$ + $Ferrocytochrome b5_{}$ + 2 $H_{}$ & $C_{18}H_{32}O_{2}$ + 2 $Ferrocytochrome b5_{}$ + 2 $H_{}$ & $O_{2}$ + $H_{}$ & $O_{2}$ + 2 $H_{}$ & $C_{18}H_{32}O_{2}$ + $O_{2}$ + $H_{}$ & $C_{18}H_{32}O_{2}$ + $O_{2}$ + 2 $H_{}$ & $Ferrocytochrome b5_{}$ + $O_{2}$ + $H_{}$ & 2 $Ferrocytochrome b5_{}$ + $O_{2}$ + $H_{}$ & $Ferrocytochrome b5_{}$ + $O_{2}$ + 2 $H_{}$ & 2 $Ferrocytochrome b5_{}$ + $O_{2}$ + 2 $H_{}$ & $C_{18}H_{32}O_{2}$ + $Ferrocytochrome b5_{}$ + $O_{2}$ + $H_{}$ & $C_{18}H_{32}O_{2}$ + 2 $Ferrocytochrome b5_{}$ + $O_{2}$ + $H_{}$ & $C_{18}H_{32}O_{2}$ + $Ferrocytochrome b5_{}$ + $O_{2}$ + 2 $H_{}$ & $C_{18}H_{32}O_{2}$ + 2 $Ferrocytochrome b5_{}$ + $O_{2}$ + 2 $H_{}$  \\
\hline$C_{18}H_{30}O_{2}$ & 0.00243 & 1.21429 & 1.41860 & 0.12260 & 0.23990 & 0.90698 & 0.02613 & 1.11364 & 1.31111 & 0.14319 & 0.25754 & 0.99513 & 0.99029 & 0.00365 & 0.00485 & 1.20903 & 1.41299 & 1.20379 & 1.40741 & 0.12365 & 0.24081 & 0.12470 & 0.24171 & 0.90255 & 0.89815 & 0.02728 & 0.02844 & 1.10884 & 1.30599 & 1.10407 & 1.30088 & 0.14420 & 0.25840 & 0.14520 & 0.25926 \\
$H_{2}O_{}$ & 0.94340 & 5.09091 & 6.62500 & 1.15207 & 1.35135 & 0.37500 & 0.94595 & 2.66667 & 4.07692 & 1.14537 & 1.33621 & 0.84615 & 0.71429 & 0.94353 & 0.94366 & 4.82609 & 6.39394 & 4.58333 & 6.17647 & 1.15172 & 1.35056 & 1.15138 & 1.34978 & 0.33333 & 0.29412 & 0.94607 & 0.94619 & 2.58140 & 3.98113 & 2.50000 & 3.88889 & 1.14505 & 1.33548 & 1.14474 & 1.33476 \\
2 $H_{2}O_{}$ & 0.88991 & 3.64706 & 5.09091 & 1.09417 & 1.28947 & 0.09091 & 0.89474 & 1.92593 & 3.18750 & 1.09013 & 1.27731 & 0.92000 & 0.84615 & 0.89016 & 0.89041 & 3.51429 & 4.95556 & 3.38889 & 4.82609 & 1.09396 & 1.28884 & 1.09375 & 1.28821 & 0.06667 & 0.04348 & 0.89497 & 0.89520 & 1.87273 & 3.12308 & 1.82143 & 3.06061 & 1.08994 & 1.27673 & 1.08974 & 1.27615 \\
$Ferricytochrome b5_{}$ & 1.21327 & 0.02778 & 0.01852 & 0.95499 & 0.93338 & 4.00000 & 1.20362 & 0.51389 & 0.41111 & 0.95698 & 0.93627 & 9.18182 & 8.50000 & 1.21277 & 1.21226 & 0.07407 & 0.05018 & 0.11616 & 0.07986 & 0.95509 & 0.93353 & 0.95520 & 0.93368 & 3.90323 & 3.81250 & 1.20316 & 1.20270 & 0.52575 & 0.42266 & 0.53704 & 0.43376 & 0.95708 & 0.93641 & 0.95717 & 0.93654 \\
2 $Ferricytochrome b5_{}$ & 1.41667 & 0.01852 & 0.01389 & 0.93338 & 0.91273 & 5.50000 & 1.39823 & 0.41111 & 0.34259 & 0.93627 & 0.91643 & 9.57143 & 9.18182 & 1.41570 & 1.41475 & 0.05018 & 0.03794 & 0.07986 & 0.06085 & 0.93353 & 0.91293 & 0.93368 & 0.91312 & 5.39024 & 5.28571 & 1.39735 & 1.39648 & 0.42266 & 0.35337 & 0.43376 & 0.36380 & 0.93641 & 0.91661 & 0.93654 & 0.91678 \\
\hline
\et}
\ec
\bc
\scalebox{0.19}{\bt{|c||ccccccccccccccccc|}
\hline & $Ferrocytochrome b5_{}$ & 2 $Ferrocytochrome b5_{}$ & $O_{2}$ & $Ferrocytochrome b5_{}$ + $O_{2}$ & 2 $Ferrocytochrome b5_{}$ + $O_{2}$ & $H_{}$ & 2 $H_{}$ & $Ferrocytochrome b5_{}$ + $H_{}$ & 2 $Ferrocytochrome b5_{}$ + $H_{}$ & $Ferrocytochrome b5_{}$ + 2 $H_{}$ & 2 $Ferrocytochrome b5_{}$ + 2 $H_{}$ & $O_{2}$ + $H_{}$ & $O_{2}$ + 2 $H_{}$ & $Ferrocytochrome b5_{}$ + $O_{2}$ + $H_{}$ & 2 $Ferrocytochrome b5_{}$ + $O_{2}$ + $H_{}$ & $Ferrocytochrome b5_{}$ + $O_{2}$ + 2 $H_{}$ & 2 $Ferrocytochrome b5_{}$ + $O_{2}$ + 2 $H_{}$  \\
\hline$H_{2}O_{}$ & 5.09091 & 6.62500 & 0.37500 & 2.66667 & 4.07692 & 0.84615 & 0.71429 & 4.82609 & 6.39394 & 4.58333 & 6.17647 & 0.33333 & 0.29412 & 2.58140 & 3.98113 & 2.50000 & 3.88889 \\
2 $H_{2}O_{}$ & 3.64706 & 5.09091 & 0.09091 & 1.92593 & 3.18750 & 0.92000 & 0.84615 & 3.51429 & 4.95556 & 3.38889 & 4.82609 & 0.06667 & 0.04348 & 1.87273 & 3.12308 & 1.82143 & 3.06061 \\
$Ferricytochrome b5_{}$ & 0.02778 & 0.01852 & 4.00000 & 0.51389 & 0.41111 & 9.18182 & 8.50000 & 0.07407 & 0.05018 & 0.11616 & 0.07986 & 3.90323 & 3.81250 & 0.52575 & 0.42266 & 0.53704 & 0.43376 \\
2 $Ferricytochrome b5_{}$ & 0.01852 & 0.01389 & 5.50000 & 0.41111 & 0.34259 & 9.57143 & 9.18182 & 0.05018 & 0.03794 & 0.07986 & 0.06085 & 5.39024 & 5.28571 & 0.42266 & 0.35337 & 0.43376 & 0.36380 \\
\hline
\et}
\ec
\bc
\scalebox{1.}{\bt{|c||ccccc|}
\hline & $O_{2}$ & $H_{}$ & 2 $H_{}$ & $O_{2}$ + $H_{}$ & $O_{2}$ + 2 $H_{}$  \\
\hline$H_{2}O_{}$ & 0.37500 & 0.84615 & 0.71429 & 0.33333 & 0.29412 \\
2 $H_{2}O_{}$ & 0.09091 & 0.92000 & 0.84615 & 0.06667 & 0.04348 \\
\hline
\et}
\ec
\bc
\scalebox{.55}{\begin{tikzpicture}
\node[draw, red] (S1) at (0, 0) {  $\underbrace{\textrm{Crepenynate}}_{C_{18}H_{30}O_{2}}$ };
\node (plus1) [right=.1 of S1] {+};
\node[draw, blue] (S2) [right=.1 of plus1] {  $2 \underbrace{\textrm{Ferricytochrome b5}}_{Ferricytochrome b5_{}}$ };
\node (plus2) [right=.1 of S2] {+};
\node[draw, brown] (S3) [right=.1 of plus2] {  $2 \underbrace{\textrm{H2O}}_{H_{2}O_{}}$ };
\node (A) [right=0.1 of S3] {$\Leftarrow$};
\node[draw, red] (P1) [right=0.1 of A] {  $\underbrace{\textrm{Linoleate}}_{C_{18}H_{32}O_{2}}$ };
\node (plus4) [right=.1 of P1] {+};
\node[draw, blue] (P2) [right=.1 of plus4] {  $2 \underbrace{\textrm{Ferrocytochrome b5}}_{Ferrocytochrome b5_{}}$ };
\node (plus5) [right=.1 of P2] {+};
\node[draw, brown] (P3) [right=.1 of plus5] {  $\underbrace{\textrm{Oxygen}}_{O_{2}}$ };
\node (plus6) [right=.1 of P3] {+};
\node[draw, brown] (P4) [right=.1 of plus6] {  $2 \underbrace{\textrm{H+}}_{H_{}}$ };
\node[draw, red] (V1) [below=3 of S1] {0.0024330900243309003};
\draw[red] (S1) -- (V1);
\draw[red] (P1) -- ($ (P1) - (0, 3) $);
\draw[red] ($ (P1) - (0, 3) $) -- (V1);
\node[draw, blue] (V2) [below=2 of S2] {0.013888888888888888};
\draw[blue] (S2) -- (V2);
\draw[blue] (P2) -- ($ (P2) - (0, 2) $);
\draw[blue] ($ (P2) - (0, 2) $) -- (V2);
\node[draw, brown] (V3) [below=1 of S3] {0.04347826086956522};
\draw[brown] (S3) -- (V3);
\draw[brown] (P3) -- ($ (P3) - (0, 1) $);
\draw[brown] ($ (P3) - (0, 1) $) -- (V3);
\draw[brown] (P4) -- ($ (P4) - (0, 1) $);
\draw[brown] ($ (P4) - (0, 1) $) -- (V3);
\end{tikzpicture}}

{\bf This reaction is not balanced!} (0.000612 points)

\bt{c|c}
 Missing in Right-HS & Missing in Left-HS \\
\hline
 $Ferrocytochrome b5_{2}$ &  $Ferricytochrome b5_{2}$ \\
\hline
\et
\ec
\end{framed}
\caption{Reaction with compound that cannot be reconstructed as combination of chemical elements.}\label{sc5}
\end{schemea}


\section*{Funding}

This work was supported by {\it Coordenação de Aperfeiçoamento de Pessoal de Nível Superior} (CAPES) and {\it Conselho Nacional de Desenvolvimento Científico e Tecnológico} (CNPq) through grant Universal No 402487/2023-0 and grant Bolsas de Produtividade em Pesquisa No 305872/2022-2.

\section*{Acknowledgments}

J.~A.~Pellizaro would like to thank the financial support of the {\it Coordenação de Aperfeiçoamento de Pessoal de Nível Superior} – Brasil (CAPES). D. Gamermann acknowledges CNPq for financial support through grant Universal No 402487/2023-0.


\bibliographystyle{unsrt}
\bibliography{surpriser}

\end{document}